\documentclass[12pt]{article}

\usepackage{amsmath}
\usepackage{indentfirst}
\usepackage{amsfonts,amssymb}
\usepackage{mathrsfs}
\usepackage{CJK,CJKnumb}
\usepackage{color}
\usepackage{amsthm}
\usepackage{graphicx}
\usepackage{enumerate}
\usepackage{epstopdf}

\numberwithin{equation}{section}

\def\sF{{\mathscr F}}

\def\sB{{\mathscr B}}
\def\sX{{\mathscr X}}

\def\sS{{\mathscr S}}
\def\sC{{\mathscr C}}
\def\sR{{\mathscr R}}

\begin{document}

\title{  { On evaluation of joint risk for non-negative \\ multivariate risks under dependence uncertainty }
         \thanks{ Supported in part by the National Natural Science Foundation of China (Nos: 12271415, 12001411)
                             and the Fundamental Research Funds for the Central Universities of China (WUT: 2021IVB024).}\\
         \footnotetext{1 School of Mathematics and Statistics, Wuhan University, Wuhan, Hubei 430072, People's Republic of China,
                         E-mail: shuogong@whu.edu.cn }
         \footnotetext{2 School of Mathematics and Statistics, Wuhan University, Wuhan, Hubei 430072, People's Republic of China,
                         E-mail: yjhu.math@whu.edu.cn }
         \footnotetext{3 School of Mathematics and Statistics, Wuhan University of Technology, Wuhan, Hubei 430070, People's Republic of China,
                         E-mail: lxwei@whut.edu.cn }
      }

\vspace{0.1cm}

\author{ Shuo Gong$^{1}$, \quad Yijun Hu$^{2}$, \quad Linxiao Wei$^{3}$ \\
       }

\vspace{0.1cm}

\date{\today}

\maketitle

\noindent{\bf Abstract:} \quad  In this paper, we propose a novel axiomatic approach to evaluating the joint risk of multiple insurance risks
under dependence uncertainty. Motivated by both the theory of expected utility and the Cobb-Dauglas utility function,
we establish a joint risk measure for non-negative multivariate risks, which we refer to as a scalar distortion joint risk measure.
After having studied its fundamental properties, we provide an axiomatic characterization of it by proposing a set of new axioms.
The most novel axiom is the component-wise positive homogeneity.
Then, based on the resulting distortion joint risk measures, we also propose a new class of vector-valued distortion joint risk measures for
non-negative multivariate risks.
Finally, we make comparisons with some  vector-valued multivariate risk measures known in the literature,
such as multivariate lower-orthant value at risk,  multivariate upper-orthant conditional-tail-expectation, multivariate tail conditional expectation
and multivariate tail distortion risk measures. It turns out that those vector-valued multivariate risk measures have forms of
vector-valued distortion joint risk measures, respectively.
This paper mainly gives some theoretical results about the evaluation of joint risk under dependence uncertainty,
and it is expected to be helpful for measuring joint risk.

\vspace{0.2cm}


\vspace{0.2cm}


\newpage

\section{Introduction}\label{sec:1}

Nowadays, risk measures are widely used in both insurance and finance, such as insurance pricing and regulatory capital calculation, etc.
Since the seminal work of Artzner et al. (1999), classical risk measures are defined on univariate risks, i.e. on random variables defined on
some measurable space $(\Omega, \sF)$, for instance, see F\"{o}llmer and Schied (2002), Frittelli and Rosazza Gianin (2002).
In practice, these classical risk measures usually require the acquirement of accurate distribution functions of the random variables.
Mathematically, it equivalently requires an accurate probability measure
on $(\Omega, \sF),$ which is sometimes known as a reference measure or a scenario in the literature.
For instance, value at risk (VaR) and expected shortfall (ES) are the cases, which are two standard risk measures popularly used in practice.
It is also well known that the insurance premium principles have a close connection with risk measures.
For instance, Wang et al. (1997) established a significant axiomatic approach
to insurance pricing for univariate insurance risks,
which is now known as distortion premium principles in the insurance literature.
For more about risk measures for random variables, we refer to F\"{o}llmer and Schied (2016).

\vspace{0.3cm}

From the practical perspective, it is usually difficult to accurately capture the \textit{true}
distribution functions of random variables, because the distribution functions usually have to be estimated from dada (samples) or
statistical hypothesis (simulation).
Therefore, model uncertainty problem naturally arises. Model uncertainty issue could stem from Knightian uncertainty; for instance,
see  F\"{o}llmer and Schied (2016, Page 5 and Section 4.2).
Model uncertainty could have different disguises in different situations. However, it is worth mentioning that there are two natural and important
approaches to deal with model uncertainty.
One approach is to consider random variables on some measurable space $(\Omega, \sF)$ without assuming a reference measure (scenario) on it;
for instance, see F\"{o}llmer and Schied (2016, Sections 4.2 and 4.7). Such models are also called \textit{model-free} models; for instance,
see Artzner et al. (1999).
Another approach is to consider multiple scenarios on $(\Omega, \sF)$; for instance, see Kou and Peng (2016, Section 3), Wang and Ziegel (2021),
Fadina et al. (2023) and the references therein.
In the present paper, we will also take into account the model uncertainty issue. Moreover, we will follow the way of considering
\textit{model-free} models.

\vspace{0.3cm}

Besides to deal with stand-alone insurance risks separately, it is also often for one to deal with multiple insurance risks as a whole.
For instance, Yuen and Guo (2001) studied ruin probabilities for two types of correlated claims.
Yuen et al. (2002) studied the ruin probability for the correlated aggregate claims.
There also have been studies about optimal multivariate reinsurance designs for multiple insurance risks;
for instance, see Denuit and Vermandele (1998), Cai and Wei (2012), Cheung et al. (2014), Zhu et al. (2014) and the references therein.
Mathematically, the random loss of an insurance risk faced by an insurer is commonly presented by a non-negative random variable defined on
some measurable apace $(\Omega, \sF),$ which could be a claim or an aggregate of claims. Thus, the random losses
of multiple insurance risks can be presented by a non-negative multidimensional random vector ${\bf X} =(X_1, \cdots, X_d)$ defined on $(\Omega, \sF).$
To assess the risk of a non-negative random vector ${\bf X} =(X_1, \cdots, X_d),$  one traditional approach is to transform
${\bf X} =(X_1, \cdots, X_d)$ into a random variable by using some aggregation procedure, for instance, using the sum of all components
$X_i.$ Once the random vector ${\bf X}$ has been transformed into a random variable, then classical univariate risk measures can be employed
to quantify the risk. Nevertheless, it is also often for one to take into account the possible
inter-dependence structure between the components of ${\bf X},$ and thus multivariate risk measures arises.
In other words, to evaluate multiple insurance risks also requires multivariate risk measures that can capture the
inter-dependence between individual insurance risks. Generally speaking, a (scalar) multivariate risk measure is to assign a random vector
with a numerical value to quantify the risk of the random vector, and thus it is a functional from a set of random vectors to the real numbers.

\vspace{0.3cm}

Burgert and R\"{u}schendorf (2006) axiomatically introduced and characterized the (scalar) multivariate coherent and convex risk measures
for random vectors ${\bf X} =(X_1, \cdots, X_d)$ defined on some probability space.
Guillen et al. (2018) established an elegant multivariate distortion risk measure for non-negative random vectors defined on some probability space.
For more related studies about (scalar) multivariate risk measures,  we refer to  R\"{u}schendorf (2006, 2013), Ekeland and Schachermayer (2011),
Ekeland et al. (2012), Wei and Hu (2014) and the references therein, just name a few.
As pointed by Burgert and R\"{u}schendorf (2006),
a multivariate risk measure is to measure not only the risk of the components (i.e.  marginals) of a random vector ${\bf X}$  separately,
but also to measure the \textit{joint risk} of ${\bf X}$ caused by the variation of the components and their possible inter-dependence.
Thus, study on measuring the \textit{joint risk} of a random vector should be helpful for understanding how much risk the \textit{joint risk}
could contribute. In the literature, study on evaluation of \textit{joint risk} is seldom carried out.
Motivated by above consideration, in this paper, we focus on the evaluation of \textit{joint risk} of non-negative random vectors.
More precisely, we will establish and characterize a new class of risk measures to evaluate the \textit{joint risk},
which we refer to as (scalar) \textit{distortion joint risk measures}.
Interestingly, it turns out that the multivariate distortion risk measures of Guillen et al. (2018) could be kinds of distortion joint risk measures;
see the sequel for the explanation. We believe that (scalar) distortion joint risk measures are worth studying.

\vspace{0.3cm}

By the Sklar's Theorem, the inter-dependence structure of a random vector is determined by the (reference) probability measure
on the measurable space. Thus, model uncertainty can lead to inter-dependence uncertainty. In the present paper, we will make use of copula to
describe the possible inter-dependence between the components of a random vector. Taking into account the possible model uncertainty issue,
we will work on the bounded  random vectors on some measurable space $(\Omega, \sF)$, rather than the essentially bounded random vectors
on certain probability space. For a model-free model $(\Omega, \sF)$, the representations for univariate risk measures usually involve monotone
set functions on $\sF$. Inspired by this, we will also employ monotone set functions on the (multidimensional) measurable
rectangles of the product $\sigma$-algebra $\sF \times \cdots \times \sF$ to establish and characterize
the (scalar) \textit{distortion joint risk measures}.

\vspace{0.3cm}

It should be mentioned that there have been many studies about vector-valued multivariate risk measures for multiple insurance risks
in the insurance literature; for instance, see Cousin and Di Bernardino (2013, 2014), Cossette et al. (2013), H\"{u}erlimann (2014),
Di Bernardino et al. (2015), Maume-Deschamps et al. (2016), Landsman et al. (2016), Cai et al. (2017), Hermann et al. (2020), Shushi and Yao (2020)
and the references therein.
For a random vector ${\bf X} = (X_1, \cdots, X_d),$ a vector-valued multivariate risk measure is to
respectively evaluate the risk of marginals $X_i$ in the context of ${\bf X},$ in which each $X_i$ is regarded as a component of ${\bf X}$
rather than a stand-alone random variable.
For instance, Cousin and Di Bernardino (2013, 2014) studied vector-valued multivariate VaR and conditional-tail-expectation (CTE),
respectively. H\"{u}rlimann (2014) studied the properties of vector-valued VaR and CTE for Archimedean copulas. Landsman et al. (2016) studied
vector-valued multivariate tail conditional expectation (TCE) for elliptical distributions. Cai et al. (2017) studied vector-valued multivariate
tail distortion risk measures. In the present paper, we will also introduce and study a type of vector-valued distortion joint risk measures for
non-negative random vectors.
For multivariate risks, there also have been a lot of set-valued risk measures and systemic risk measures available
in the literature. Since the focus of this paper will be the scalar and vector-valued joint risk measures, we temporarily do not intend to
make a literature review about set-valued risk measures and systemic risk measures.

\vspace{0.3cm}

In this paper, we focus on the scalar and vector-valued joint risk measures for non-negative random vectors (i.e. for multiple insurance risks)
under model uncertainty (i.e. under inter-dependence uncertainty).
Motivated by both the theory of expected utility and the Cobb-Dauglas utility function, we establish a joint risk measure  
to evaluate the \textit{joint risk} of non-negative random vectors, which we refer to as a (scalar) distortion joint risk measure.
After having studied its fundamental properties, we provide an axiomatic characterization of it by proposing a set of new axioms.
The most novel axiom is the component-wise positive homogeneity, see Axiom (A1) below.
Then, based on the resulting distortion joint risk measures, we also propose a new class of vector-valued distortion joint risk measures for
non-negative random vectors.
Finally, we make comparisons with some known vector-valued multivariate risk measures,
such as vector-valued multivariate lower-orthant VaR and upper-orthant CTE, vector-valued multivariate TCE
and vector-valued multivariate tail distortion risk measures. It turns out that those vector-valued multivariate risk measures have forms of
vector-valued distortion joint risk measures, respectively.
This paper mainly gives some theoretical results about the evaluation of joint risk under dependence uncertainty,
and it is expected to be helpful for measuring joint risk.

\vspace{0.3cm}

It should be mentioned that the most relevant literature is the one of Guillen et al. (2018).
However, the difference between the study of Guillen et al. (2018) and that of this paper is significant.
First, the angle of this paper is different from that of Guillen et al. (2018).
Inspired by both the expected utility theory and the Cobb-Dauglas utility function, we establish
a distortion joint risk measure from the perspective of \textit{joint risk}. This viewpoint of joint risk measures might be helpful to
provide a plausible interpretation for the multivariate distortion risk measures of Guillen et al. (2018).
For more details about the connection with the multivariate distortion
risk measures of Guillen et al. (2018), see the discussions after Definition 3.1 below.
Second, we provide an axiomatic characterization of distortion joint risk measures by proposing
a set of new axioms, whereas axiomatic characterization is not discussed in Guillen et al. (2018).
The most novel component-wise positive homogeneity also indicates that what we measure is the \textit{joint risk} of multiple risks;
see the comments on the component-wise positive homogeneity after Remark 3.1 below.
Third, we also reveal a connection of the (scalar) distortion joint risk measures with some vector-valued multivariate risk measures known
in the literature; see Subsection 3.3 and Section 4 below for details.
Taking the above considerations into account, this study can be viewed as a meaningful complement to the study of Guillen et al. (2018).

\vspace{0.3cm}

The rest of this paper is organized as follows. In Section 2, we provide preliminaries.
 Section 3 is devoted to the statements of the main results of this paper.
In the first subsection, we introduce some new axioms for joint risk measures, as well as financial interpretations.
In the second subsection, we establish scalar distortion joint risk measures, and axiomatically characterize them.
In the final subsection, we construct vector-valued distortion
joint risk measures, and study their fundamental properties. In Section 4, comparisons with some known vector-valued
risk measures are made. Concluding remarks are summarized in Section 5.
In the appendix, we provide the proofs of all main results of this paper. For the self-contained purpose,
we also address relevant issues in the appendix.

\section{Preliminaries}

Let $(\Omega, \sF)$ be a measurable space, and $P$ a fixed probability measure on it, acting as a reference measure,
which is also known as a scenario in some literature, for instance, see Kou and Peng (2016, Section 3), Wang and Ziegel (2021)
and Fadina et al. (2023).
We denote by $\sX$ the linear space of all bounded measurable functions (i.e. random variables) on $(\Omega, \sF)$
equipped with the supremum norm $\|\cdot\|$, and by $\sX_+$ the subset of $\sX$ consisting of those elements which are non-negative.
To be consistent with the literature on risk measures under model uncertainty,
we would like to concentrate on $\sX$ rather than $L^\infty(\Omega, \sF, P)$
of essentially bounded random variables on $(\Omega, \sF, P)$,
for instance, see F\"{o}llmer and Schied (2016, Sections 4.2 and 4.7), Kou and Peng (2016, Section 3) or Wang and Ziegel (2021).
Let $d \geq 1$ be a fixed integer. We denote by $\sX^d$ the space of all $d$-dimensional random vectors on $(\Omega, \sF)$
with each component (i.e. marginal) belonging to $\sX$, and by $\sX^d_+$ the set of all $d$-dimensional random vectors
on $(\Omega, \sF)$ with each component belonging to $\sX_+.$

\vspace{0.3cm}

In the sequel, we will focus on joint risk measures on $\sX^d_+.$
A random variable $X \in \sX_+$ represents the random loss of an insurance risk faced by an insurer, for instance,
it can be a claim or an aggregate of claims.
Thus, a random vector ${\bf X} = (X_1, \cdots, X_d) \in \sX^d_+$ represents the random losses of $d$ insurance risks,
in which $X_i$ represents the random loss of the $i$th insurance risk.
Although we concentrate on multiple insurance risks (i.e. on $\sX^d_+$), the resulting distortion joint risk measures are also applicable
for multiple financial risks. Therefore, in the Appendix A.3, we will briefly make an address about general random vectors
(i.e. $\sX^d$),
in which a random variable $X \in \sX$ represents the random loss of a financial position.

\vspace{0.3cm}

Generally speaking, a risk measure can be defined as any functional on $\sX$ or $\sX_+$.
Suppose that $\mu$ is a normalized and monotone set function from $\sF$ to $[0,1]$, that is, $\mu(\emptyset) = 0,$  $\mu(\Omega)=1$ and
$\mu(A) \leq \mu(B)$ for any $A, B \in \sF$ with $A \subseteq B,$ where $\emptyset$ stands for the empty set.
Then for $X\in \sX$, the Choquet integral of $X$ with respect to $\mu$
is defined as
$$
 \int X d\mu := \int_{-\infty}^0[\mu(X> x)-1]dx+\int_0^{\infty}\mu(X> x)dx.
$$

\vspace{0.3cm}

It is well known that the distortion risk measure has a close relationship with the Choquet integral. Given a non-decreasing function
$g : [0,1] \rightarrow [0,1]$ with $g(0)=0$ and $g(1)=1$, the distortion risk measure $\rho_g : \sX \rightarrow \mathbb{R}$ is defined
as
\begin{align*}
  \rho_g(X):= \int_{-\infty}^0 \left[ g(P(X > x))-1 \right]dx + \int_0^{+\infty} g(P(X > x)) dx, \quad X \in \sX.
\end{align*}
Such a function $g$ is called a distortion function, and the composition $g \circ P$ of $g$ and $P$ is call a distorted probability.
It is well known that the distortion risk measure includes VaR and ES (also known as
conditional value at risk (CVaR)), for instance, see F\"{o}llmer and Schied (2016) or Wang and Ziegel (2021).

\vspace{0.3cm}

Let $\sS$ be the set of all $d$-dimensional measurable rectangles of the product $\sigma$-algebra $\sF^d := \sF \times \cdots \times \sF,$
that is $\sS := \{ A_1 \times \cdots \times A_d : A_i \in \sF, 1\leq i\leq d \}$. 
A mapping  $\mu: \sS \rightarrow [0, +\infty) $ is called a set function on $\sS$, if  $\mu({\bf \emptyset } ) = 0, $
where $ {\bf \emptyset } := A_1 \times \cdots \times A_d \in \sS $ with some $A_i$ being empty set. 
Given a set function $\mu$ on $\sS$, for ${\bf A} = A_1 \times \cdots \times A_d \in \sS,$ we also write
$\mu ( A_1 \times \cdots \times A_d)$ for $\mu({\bf A})$ if the emphasis is on $A_i, 1 \leq i \leq d.$
A set function $\mu$ on $\sS$ is called normalized if $\mu (\Omega \times \cdots \times \Omega) = 1$,
and monotone if $ \mu ({\bf A}) \leq \mu ({\bf B})$
for all ${\bf A} = A_1 \times \cdots \times A_d,$ ${\bf B} = B_1 \times \cdots \times B_d \in \sS $ with $A_i \subseteq B_i, 1 \leq i \leq d.$
For a sequence of sets $ \{ {\bf D}^{(n)} = D^{(n)}_1 \times \cdots \times D^{(n)}_d \in \sS; n \geq 1 \}$
and ${\bf D} = D_1 \times \cdots \times D_d \in \sS,$
we say that the sequence of sets $ \{ {\bf D}^{(n)}; n \geq 1 \}$ increases to ${\bf D}$,
if for any $n \geq 1,$
$ D^{(n)}_i \subseteq D^{(n+1)}_i \subseteq D_i$  and $  \underset{n \rightarrow +\infty}{\lim} D^{(n)}_i = D_i$
for each $1 \leq i \leq d.$ In this situation, we denote $ {\bf D}^{(n)}\uparrow {\bf D}.$
A set function $\mu$ on $\sS$ is called continuous from below if $\underset{n \rightarrow +\infty}{\lim} \mu({\bf D}^{(n)})=\mu({\bf D})$
for any sequence of sets $ \{ {\bf D}^{(n)} \in \sS$; $n \geq 1 \}$ and ${\bf D} \in \sS$ with ${\bf D}^{(n)} \uparrow  {\bf D}.$

\vspace{0.3cm}
Similar to $d$-variate functions, we can define the increment of any set function $\mu$ on $\sS.$
To be precise, for any ${\bf A} = A_1 \times \cdots \times A_d, {\bf B} = B_1 \times \cdots \times B_d \in \sS $,
the increment of $\mu$ from ${\bf A}$ to ${\bf B}$ is defined by
\begin{align*}
 \Delta^{{\bf B}}_{{\bf A}} \mu := \Delta_{{\bf B},{\bf A}} \mu
                                := \Delta^{(d)}_{B_d, A_d} \Delta^{(d-1)}_{B_{d-1}, A_{d-1}} \cdots \Delta^{(1)}_{B_1, A_1} \mu,
\end{align*}
where for any set function $\nu : \sS \rightarrow [0, +\infty),$
\begin{align*}
\Delta^{(i)}_{B_i, A_i} \nu(D_1, \cdots, D_d)
  & := \nu(D_1 \times \cdots \times D_{i-1} \times B_i \times D_{i+1}\times \cdots\times D_d)\\
  & \quad -  \nu(D_1\times \cdots \times D_{i-1}\times A_i\times D_{i+1}\times \cdots\times D_d).
\end{align*}
A set function $\mu$ on $\sS$ is called $d$-monotone if  $ \Delta_{{\bf B}, {\bf A}} \mu \geq 0$
for all ${\bf A} = A_1 \times \cdots \times A_d, {\bf B} = B_1 \times \cdots \times B_d \in \sS $ with $A_i \subseteq B_i, 1 \leq i \leq d.$
Note that the $d$-monotonicity of a set function implies the monotonicity, but the converse assertion may not be true.

\vspace{0.3cm}

For ${\bf X}= (X_1,\cdots,X_d) \in \sX^d$, denote by
\begin{align*}
F_{\bf X}({\bf x}):= P({\bf X} \leq {\bf x}):= P(X_1\leq x_1,\cdots,X_d\leq x_d), \quad {\bf x}=(x_1,\cdots,x_d) \in \mathbb{R}^d,
\end{align*}
and
\begin{align*}
S_{\bf X}({\bf x}):= P({\bf X}> {\bf x}):= P(X_1>x_1,\cdots,X_d>x_d), \quad {\bf x}= (x_1,\cdots,x_d) \in \mathbb{R}^d,
\end{align*}
the joint distribution and joint survival functions of ${\bf X}$ under  $P$, respectively.
Similarly, given a $d$-monotone and continuous from below set function $\mu: \sS \rightarrow [0, +\infty)$, the distribution function of ${\bf X}$
with respect to $\mu$, denoted by $F_{\mu,{\bf X}}$, is defined by
\begin{align*}
  F_{\mu, {\bf X}}({\bf x}) := \mu (\{X_1 \leq x_1\} \times \cdots \times \{X_d \leq x_d\}), \quad {\bf x}=(x_1,\cdots,x_d) \in \mathbb{R}^d,
\end{align*}
and the survival function of ${\bf X}$ with respect to $\mu$, denoted by $S_{\mu,{\bf X}}$, is defined by
\begin{align*}
  S_{\mu, {\bf X}}({\bf x}) := \mu (\{X_1 > x_1\} \times \cdots \times \{X_d > x_d \}), \quad {\bf x}=(x_1,\cdots,x_d) \in \mathbb{R}^d.
\end{align*}

\vspace{0.3cm}

Next, we recall the definitions of sub-copula and copula. For more details, we refer to Nelsen (2006).

\vspace{0.3cm}

\noindent\noindent{\bf Definition 2.1}\ \ Denote by $\sB([0,1])$ the Borel algebra of Borel subsets of $[0,1]$.
Given $A_i \in \sB([0,1])$, $1 \leq i \leq d,$ with $ 0,1 \in A_i$. A function $C: A_1\times \cdots \times A_d \rightarrow [0,1]$ is called
a sub-copula if it satisfies the following three properties:
\begin{enumerate}
  \item[(i)] $C(x_1,\cdots,x_d)=0$, if there is an $i \in \{1,\cdots,d\}$ such that $x_i=0$.
  \item[(ii)] $C(1,\cdots,1,x_i,1,\cdots,1)= x_i$, $1\leq i\leq d$.
  \item[(iii)] $C$ is $d$-monotone, that is, $ \Delta_{{\bf b}, {\bf a}} C \geq 0$ for any ${\bf a},{\bf b} \in A_1\times \cdots \times A_d$
                    with ${\bf a} \leq {\bf b}$.
\end{enumerate}
Furthermore, if $A_i=[0,1]$, $1\leq i\leq d,$  then the sub-copula is called a copula. Denote by $\sC$ the set of all copulas on $[0,1]^d$.
Note that in the above definition, $ \Delta_{{\bf b}, {\bf a}} C $ stands for the increment of $C$ on the interval $[{\bf a}, {\bf b}],$
see also Appendix A.2 for its definition.

\vspace{0.3cm}

By the Sklar's Theorem,
for any random vector ${\bf X}$ $=$ $(X_1, \cdots, X_d) \in \sX^d,$ there exists a copula $C$ such that
\begin{align*}
  F_{\bf X}(x_1,\cdots,x_d)= C(F_{X_1}(x_1),\cdots,F_{X_d}(x_d)) \quad \mbox{for\ every} \  (x_1,\cdots,x_d)\in \mathbb{R}^d,
\end{align*}
where $F_{X_i}(x_i):= P(X_i \leq x_i), x_i \in \mathbb{R}$, is the distribution function of ${X_i}$ under $P$, $1\leq i\leq d.$
In this case, we say that $C$ is a copula for ${\bf X},$ or that ${\bf X}$ has a copula $C.$
Furthermore, if ${\bf X}$ is a continuous random vector, then the copula for ${\bf X}$ is unique.
Moreover, for the survival function $S_{\bf X}$, there exists another copula $\widehat{C}$ such that
\begin{align*}
  S_{\bf X}(x_1,\cdots,x_d)= \widehat{C}(S_{X_1}(x_1),\cdots,S_{X_d}(x_d)) \quad \mbox{for\ every} \  (x_1,\cdots,x_d)\in \mathbb{R}^d,
\end{align*}
where $S_{X_i}:= 1- F_{X_i}$ is the survival function of ${X_i}$ under $P$, $1\leq i\leq d.$
Such a copula $\widehat{C}$ is called the survival copula for ${\bf X}$.
Note that given a copula $C$ for ${\bf X},$ the survival copula $\widehat{C}$ can be given by
\begin{align}
 \widehat{C} (u_1,\cdots,u_d)
    & := \Delta_{(1, \cdots, 1), (1-u_1, \cdots, 1-u_d)} C \nonumber\\
    & \ = 1 - [(1-u_1) + \cdots + (1-u_d)] \nonumber\\
    & \ \quad + [C(1-u_1, 1-u_2, 1, \cdots, 1) + \cdots + C(1,\cdots,1,1-u_{d-1},1-u_d)] \nonumber\\
    & \ \quad - \cdots + (-1)^d C(1-u_1,\cdots,1-u_d), \label{2002}
\end{align}
see Nelsen (2006, page 32) for the case of $d=2$. In this situation, we also say that $\widehat{C}$ is associated with $C.$
For more details, we refer to Nelsen (2006).

\vspace{0.3cm}

Given a normalized, $d$-monotone and continuous from below set function $\mu$ on $\sS$,
we denote by $L_{\mu, {\bf X}}$ the Lebesque-Stieltjes (L-S) measure induced by the survival function $S_{\mu,{\bf X}}$ of a random vector
${\bf X} \in \sX^d$ with respect to $\mu,$ which is a probability measure on the Borel algebra $\sB(\mathbb{R}^d)$.
For more details about $L_{\mu, {\bf X}}$, see Appendix A.2 or Denneberg (1994).

\vspace{0.3cm}

For convenience, we introduce more notations.  Operations on $\sX^d$ are understood in component-wise sense.
Given a functional $\Gamma : \sX^d \rightarrow \mathbb{R}$, for ${\bf X} = (X_1, \cdots, X_d) \in \sX^d$, we also write
$\Gamma(X_1, \cdots, X_d)$ for $\Gamma({\bf X})$ if the emphasis is on $X_i, 1 \leq i \leq d.$
 $\Delta^{{\bf Y}}_{{\bf X}} \Gamma $ stands for the increment of $\Gamma$ from ${\bf X}$ to ${\bf Y}$, see Appendix A.2 for its definition.
For $a,b \in \mathbb{R}$, $a \wedge b$ stands for $\min(a,b)$.
${\bf 0}:= (0,\cdots,0)$, ${\bf 1}:= (1,\cdots,1) \in \mathbb{R}^d$. For $1\leq i\leq d,$ ${\bf e}_i := (0,\cdots,0,1,0,\cdots,0)$, where $1$ occupies
the $i$th coordinate, and $\bar{{\bf e}}_i := (1, \cdots, 1, 0, 1 \cdots, 1)$, where $0$ occupies the $i$th coordinate.
For a non-empty set $A$, $1_A$ stands for the indicator function of $A$, while $1_{\emptyset} := 0$ with convention.
For a mapping $h$, Ran($h$) stands for the range of $h$.
$\mathbb{R}_+ := \left[0,+\infty\right).$ $ \mathbb{R}^d_+ := [0, +\infty)^d, d \geq 1.$

\vspace{0.3cm}

Next, we recall the definition of comonotonicity,  For more details about comonotonicity,
we refer to Denneberg (1994) or F\"{o}llmer and Schied (2016).

\vspace{0.3cm}

\noindent{\bf Definition 2.2}\ \ Two random variables $X, Y \in \sX$ are called comonotone,
if $(X(\omega_1)-X(\omega_2))(Y(\omega_1)-Y(\omega_2)) \geq 0$  for every $\omega_1,\omega_2 \in \Omega$.
For any ${\bf X} = (X_1,\cdots,X_d)$, ${\bf Y} = (Y_1,\cdots,Y_d) \in \sX^d$, ${\bf X}$ and ${\bf Y}$ are called comonotone
if $X_i$ and $Y_i$ are comonotone for each $1 \leq i \leq d$.

\vspace{0.3cm}

\noindent{\bf Remark 2.1}\ \ Two random variables $X$ and $Y$ are comonotone if and only if there exist continuous and non-decreasing functions
$h_1, h_2$ on $\mathbb{R}$ such that $h_1(u)+h_2(u)=u$, $u\in \mathbb{R}$, and $X = h_1(X+Y), Y = h_2(X+Y)$;
or equivalently, there exist a random variable $Z$ and non-decreasing functions $h_1, h_2$ on $\mathbb{R}$ such that $X = h_1(Z), Y = h_2(Z).$
For instance, see Denneberg (1994, Proposition 4.5) or F\"{o}llmer and Schied (2016).

\vspace{0.3cm}

We end this section with two lemmas, which are crucial for our study. For the self-contained purpose,
their brief proofs will be provided in Appendix A.1.

\vspace{0.3cm}

\noindent{\bf Lemma 2.1}\ \ Let a random vector ${\bf X}=(X_1,\cdots,X_d) \in \sX^d_+$ have a copula $C.$

(1) If functions $g_1,\cdots,g_d$ are continuous and non-decreasing, then $C$ is also a copula for the random vector $(g_1(X_1),\cdots,g_d(X_d)).$

(2) For any real numbers $a_1,\cdots,a_d$, $C$ is a copula for the random vector $(1_{\{X_1>a_1\}},\cdots,$ $1_{\{X_d>a_d\}}).$

\vspace{0.3cm}

Note that if the random vector ${\bf X}=(X_1,\cdots,X_d)$ is continuous, and $g_1,\cdots,g_d$ are strictly increasing on
Ran$(X_1)$, $\cdots$, Ran$(X_d)$, respectively, then Lemma 2.1(1) is shown by Theorem 2.4.3 of Nelsen (2006).

\vspace{0.3cm}

\noindent{\bf Lemma 2.2}\ \ Let $n_i$ be a positive integer, $1\leq i\leq d.$ Then for any integers $k_i,l_i \in \{ 1,\cdots,n_i\cdot 2^{n_i} \}$,
$1\leq i\leq d,$ and any ${\bf X} = (X_1,\cdots,X_d) \in \sX^d_+$ which has a copula $C,$

(1) the two random vectors
$$
  \left(\sum\limits_{j_1=1}^{k_1} \left( X_1 \wedge \frac{j_1+1}{2^{n_1}} - X_1 \wedge \frac{j_1}{2^{n_1}}\right), \cdots,
         \sum\limits_{j_d=1}^{k_d} \left( X_d \wedge \frac{j_d+1}{2^{n_d}} - X_d \wedge \frac{j_d}{2^{n_d}}\right)\right)
$$
and
$$
  \left(\sum\limits_{j_1=1}^{l_1} \left( X_1 \wedge \frac{j_1+1}{2^{n_1}} - X_1 \wedge \frac{j_1}{2^{n_1}}\right), \cdots,
           \sum\limits_{j_d=1}^{l_d} \left( X_d \wedge \frac{j_d+1}{2^{n_d}} - X_d \wedge \frac{j_d}{2^{n_d}}\right)\right)
$$
are comonotone.

(2) the random vectors
$$
 \left(\sum\limits_{j_1=1}^{k_1} \left( X_1 \wedge \frac{j_1+1}{2^{n_1}} - X_1 \wedge \frac{j_1}{2^{n_1}} \right), \cdots,
     \sum\limits_{j_d=1}^{k_d} \left( X_d \wedge \frac{j_d+1}{2^{n_d}} - X_d \wedge \frac{j_d}{2^{n_d}}\right)\right)
$$
and $( Y_1, \cdots, Y_d )$ are comonotone, where one of the components of $( Y_1, \cdots, Y_d )$, say $Y_l$,
equals to $( X_l \wedge \frac{k_l+2}{2^{n_l}} - X_l \wedge \frac{k_l+1}{2^{n_l}} )$,
and the others $Y_i, i\neq l$, equal to $\sum_{j_i=1}^{k_i} ( X_i \wedge \frac{j_i+1}{2^{n_i}} - X_i \wedge \frac{j_i}{2^{n_i}} )$, respectively.

(3) $C$ is a copula for the following four random vectors:
\begin{align*}
  \left( {2^{n_1}}\cdot\left( X_1\wedge \frac{k_1+1}{2^{n_1}}- X_1\wedge \frac{k_1}{2^{n_1}}\right), \cdots,
             {2^{n_d}} \cdot \left(X_d\wedge \frac{k_d+1}{2^{n_d}}-  X_d\wedge \frac{k_d}{2^{n_d}}\right) \right),
\end{align*}
\begin{align*}
  \left( {2^{n_1}}\cdot\left( X_1\wedge \frac{k_1}{2^{n_1}}- X_1\wedge \frac{k_1-1}{2^{n_1}}\right), \cdots,
                  {2^{n_d}} \cdot \left(X_d\wedge \frac{k_d}{2^{n_d}}- X_d\wedge \frac{k_d-1}{2^{n_d}}\right) \right),
\end{align*}
\begin{align*}
\left(X_1\wedge\left(n_1-\frac{1}{2^{n_1}}\right), \cdots, X_d\wedge\left(n_d-\frac{1}{2^{n_d}}\right)\right)
\end{align*}
and
\begin{align*}
  \left( 1_{\left\{X_1> \frac{k_1}{2^{n_1}}\right\}}, \cdots, 1_{\left\{X_d> \frac{k_d}{2^{n_d}}\right\}} \right).
\end{align*}

\section{ Distortion joint risk measures }

In the first subsection, we will propose some axioms (i.e. properties) for joint risk measures.
In the second subsection, we will establish (scalar) distortion joint risk measures for non-negative random vectors.
After their fundamental properties have been investigated, we axiomatically characterize them.
Based on the representations for distortion joint risk measures, the third subsection is devoted to the construction of
a vector-valued distortion joint risk measure, as well as the study of its properties.

\vspace {0.3cm}

For the sake of presenting the main results, the proofs of all main results of this section will be postponed to Appendix A.1.
Although we concentrate on joint risk measures for multiple insurance risks (i.e. for $\sX^d_+$), the joint risk measures are also applicable
for multiple financial risks. Therefore, in the Appendix A.3, we will briefly make an address about joint risk measures
for general random vectors (i.e. for $\sX^d$).

\subsection{ Axioms for joint risk measures }

In this subsection, we will introduce  some axioms (properties) for joint risk measures for non-negative random vectors.

\vspace{0.3cm}

In general, a (scalar) joint risk measure is defined as any functional $\tau: \sX^d_+  \rightarrow \mathbb{R}_+$.
A joint risk measure $\tau$  is called normalized, if $\tau({\bf 1}) = 1$.
For a random vector ${\bf X} \in \sX^d_+$, the quantity $\tau({\bf X})$ can be interpreted as either the minimal amount of liquidity requirement
caused by the joint risk of ${\bf X}$, or the minimal amount of financial resource that can cancel the joint risk of ${\bf X}$.
Here, liquidity refers to the ability of an agent to make cash payments as they become due, for instance, see Hull (2015).
This interpretation of liquidity requirement is consistent with the interpretation of capital requirement
for univariate risk measures. In fact, if $d = 1$, then from the viewpoint of regulation, the liquidity requirement
caused by the \textit{joint risk} of an one-dimensional random vector can be naturally understood as nothing else
but just the capital requirement for the random variable.

\vspace{0.3cm}

Now, we list some axioms (properties) for any joint risk measure $\tau:$ $ \sX^d_+  \rightarrow$ $ \mathbb{R}_+$,
which we will concern in the sequel.

\begin{enumerate}
  \item[(A1)] Component-wise positive homogeneity: $\tau ((c_1 X_1, \cdots, c_d X_d)) = c_1 \cdots  c_d \tau({\bf X})$
              for any $c_1, \cdots, c_d \geq 0$ and any ${\bf X}= (X_1,\cdots,X_d) \in \sX^d_+$.

  \item[(A2)] Monotonicity: $\tau({\bf X}) \leq \tau({\bf Y})$  for any ${\bf X}, {\bf Y} \in \sX^d_+$ with ${\bf X} \leq {\bf Y}$.

  \item[(A3)] Comonotone additivity: $\tau({\bf X}+{\bf Y}) = \underset{{\bf Z}}{\sum} \tau({\bf Z})$
              for any comonotone ${\bf X} = (X_1, \cdots, X_d)$ and ${\bf Y} = (Y_1, \cdots, Y_d) \in \sX^d_+$,
              where the summation $ \underset{{\bf Z}}\sum$ is summed over the $2^d$ random vectors
              ${\bf Z} = (Z_1, \cdots, Z_d)$ with $Z_i$ being either $X_i$ or $Y_i$ for each $1\leq i\leq d$.

  \item[(A4)] Continuity from below: Given ${\bf X} \in \sX^d_+$,
             $\underset{n \rightarrow +\infty}{\lim} \tau({\bf X}_n)=\tau({\bf X})$ for any sequence of random vectors
             $\{{\bf X}_n; n \geq 1\}$ with ${\bf X}_n \in \sX^d_+$, $n \geq 1$, and ${\bf X}_n \uparrow {\bf X}$.

  \item[(A5)] $d$-monotonicity: $\Delta_{\bf X}^{\bf Y} \tau \geq 0$ for any ${\bf X},$ ${\bf Y}$ $\in \sX^d_+$ with ${\bf X} \leq {\bf Y}$.

  \item[(A6)] Distribution invariance: $\tau({\bf X})=\tau({\bf Y})$ for any random vectors
                ${\bf X}$ and ${\bf Y} \in \sX^d_+$ with the same joint distribution function under the probability measure $P$.
\end{enumerate}

\vspace {0.3cm}

\noindent{\bf Remark 3.1}\ \ Both (A2) and  (A3) imply (A1).
For the self-contained purpose, a brief proof of this remark is provided in Appendix A.1.

\vspace {0.3cm}

Before we interpret above axioms in the context of finance, we would like to make two comments on the component-wise positive homogeneity (A1).
First, from Axiom (A1), we can easily conclude that $\tau({\bf X}) = 0$, whenever there is at least one component
of ${\bf X}=(X_1, \cdots, X_d)$ being zero. For simplicity and without any loss of generality, we assume that $X_d = 0$.
Then $\tau ((X_1, \cdots, X_{d-1}, 0)) = 0$. This conclusion means that in such a case, one only needs to concern the
\textit{joint risk} of the $(d-1)$-dimensional random vector $(X_1, \cdots, X_{d-1})$ rather than that of $(X_1, \cdots, X_{d-1}, 0)$, because adding
a zero random variable to the random vector $(X_1, \cdots, X_{d-1})$ undoubtedly does not change anything about the \textit{joint risk} of
$(X_1, \cdots, X_{d-1})$.
In other words, the acting \textit{joint risk} of $(X_1, \cdots, X_{d-1}, 0)$ is nothing else but just that of $(X_1, \cdots, X_{d-1})$.
This is also the main reason why we call the resulting risk measures \textit{joint risk measures} from the axiomatic point of view.
In summary, when evaluating the joint risk of a random vector, we should eliminate those zero-valued components first,
and then measure the joint risk of the remaining components using a joint risk measure on some lower dimension space $ \sX^l_+$,
$1 \leq l < d$.
We would also like to mention that the converse of above conclusion might not be true.
A sufficient condition for the converse being true is that $\tau ({\bf X}) > 0$ for any ${\bf X} = (X_1, \cdots, X_d) \in \sX^d_+$
with  $\| X_i \| > 0$ for each $1 \leq i \leq d.$
The financial meaning of this sufficient condition is this: there should have certain amount of liquidity requirement,
whenever each component of a random vector has an \textit{actual} loss, no matter what the amounts of the components'
losses are. If we had previously assumed that this sufficient condition holds as an axiom, then in the Definition 3.1 below, we should
need and only need to assume a corresponding assumption on the $d$-monotone set function $\mu$ that $\mu(A) > 0$ for any non-empty set $A \in \sS.$
Second, the component-wise positive homogeneity has essential distinction with the classic positive homogeneity (PH). The classic PH
for a multivariate risk measure states that given a multivariate risk measure $\rho:$ $\sX^d \rightarrow \mathbb{R}$,
\begin{align*}
\rho(c{\bf X}) := \rho((cX_1, \cdots, cX_d)) = c \rho({\bf X})
                                \quad \mbox{for\ any} \ {\bf X} = (X_1, \cdots, X_d) \in \sX^d \ \mbox{and\ any} \ c \geq 0,
\end{align*}
for instance, see Burgert and R\"{u}schendorf (2006), R\"{u}schendorf (2013) and Wei and Hu (2014). The classic PH reflects the impact of
change of the random vector on the corresponding capital requirement, which reveals a linear relationship between the changes of the random vector and
the capital requirement. In contract, the component-wise PH reflects the impact of change of each component of
a random vector on the corresponding liquidity requirement caused by the joint risk of the random vector, which is a linear relationship.
This characteristic has some similarity to the classic PH for univariate risk measures. For univariate risk measures, the PH reflects a linear
relationship between the changes of the random variable and capital requirement, see Artzner et al. (1999) and F\"{o}llmer and Schied (2016).
Nevertheless, from the perspective of whole random vector rather than individual components,
the component-wise PH reveals a non-linear relationship between the changes of the random vector and the liquidity requirement caused by the joint risk
of the random vector. We argue that such a non-linear relationship should be, more or less, in accordance with intuition.
This is because that there could be sort of endogenous incentive impact on the liquidity requirement regarding the performance of the joint risk,
and such a endogenous incentive impact on the liquidity requirement might not be linear.

\vspace {0.3cm}

Now, we turn to the financial interpretations of above axioms.
The component-wise PH means that a linear change of each marginal would result in a same linear change of
the corresponding liquidity requirement.
The monotonicity says that a random vector with larger losses would yield a higher liquidity requirement.
This axiom is totally similar to the classic monotonicity; for instance, see Artzner et al. (1999) and Wang et al. (1997) for univariate risk measures,
and Burgert and R\"{u}schendorf (2006), R\"{u}schendorf (2013) and Wei and Hu (2014) for multivariate risk measures.
Now we discuss the financial implication of the comonotone additivity. Given two random vectors
${\bf X}= (X_1, \cdots, X_d),$ ${\bf Y} = (Y_1, \cdots, Y_d) \in \sX^d_+$,
write ${\bf Z} = (Z_1, \cdots, Z_d)$, where $Z_i$ is either $X_i$ or $Y_i$ for each $1 \leq i \leq d$. Then ${\bf Z}$ is considered as
a sub-random-vector of the random vector ${\bf X} + {\bf Y}$ in the sense that there is a random vector ${\bf W} = (W_1, \cdots, W_d)$ with $W_i$ being
either $X_i$ or $Y_i$, $1 \leq i \leq d$, so that ${\bf X} + {\bf Y}$ $ = $ ${\bf Z} + {\bf W}$.
Hence, the comonotone additivity says that spreading joint risk of ${\bf X} + {\bf Y}$
within comonotone sub-random-vectors can not reduce the total joint risk. In fact, when $d = 1$,
then the comonotone additivity reduces to the classic comonotone additivity for univariate risk measures;
for instance, see Denneberg (1994), F\"{o}llmer and Schied (2016) and Wang et al. (1997).
In this sense, the comonotone additivity (A3)
could also be viewed as a multivariate extension of the classic univariate setting.
The axiom of continuity from below is more or less due to the technical purpose. Nevertheless, it says that
a small increment of any marginal of a random vector should not lead to a tremendous increment of the corresponding liquidity requirement.
The $d$-monotonicity is more or less a kind of technical condition, but it implies the monotonicity.
The distribution invariance means that for two random vectors, if they have the same joint distribution function under the reference probability measure $P$,
then their joint risks should be the same.  This characteristic has some similarity to the law invariance in the classic setting,
see Kusuoka (2001), F\"{o}llmer and Schied (2016), R\"{u}schendorf (2006, 2013), and Wang et al. (1997).

\subsection{Scalar distortion joint risk measures}

In this subsection, we will first establish (scalar) distortion joint risk measures. After have studying their fundamental properties,
then we provide axiomatic characterization of them.

\vspace {0.3cm}

It is well known that the theory of risk measures has a close relationship with the theory of expected utility.
Motivated by the Cobb-Douglas utility function, for a random vector ${\bf X} \in \sX^d_+,$ we define the joint risk measure of ${\bf X}$
by the expected utility with respect to the Lebesque-Stieltjes measure induced by the survival function $S_{\mu, {\bf X}}$ of ${\bf X}$
with respect to certain set function $\mu$ on $\sS,$ in which the utility function is chosen to be the Cobb-Douglas utility function with
all parameters being one. Such a consideration leads to the following definition of (scalar) distortion joint risk measures.

\vspace {0.3cm}

\noindent{\bf Definition 3.1} \ \ Let $\mu: \sS \rightarrow \mathbb{R}_+$ be a normalized, $d$-monotone and continuous from below set function.
The (scalar) distortion joint risk measure $\Gamma_{\mu}: \sX^d_+  \rightarrow \mathbb{R}_+$ is defined by
\begin{align}\label{3001}
 \Gamma_{\mu}({\bf X})
     := \int_{\mathbb{R}^d_+} x_1  \cdots  x_d L_{\mu, {\bf X}} (d x_1, \cdots, d x_d), \quad  {\bf X} \in \sX^d_+.
\end{align}
Here, we have used a subscript $\mu$ in the notation $\Gamma_{\mu}$ to indicate that the set function $\mu$ on $\sS$ is pre-specified.

\vspace {0.3cm}

Notice that the distortion joint risk measure $\Gamma_{\mu}$ as in (\ref{3001}) also has the following expression, which we will
often adopt in the sequel: for ${\bf X}=(X_1,\cdots,X_d) \in \sX^d_+,$
\begin{align}\label{3002}
 \Gamma_{\mu}({\bf X})
      = \int_0^{\infty}\cdots\int_0^{\infty} \mu(\{X_1>x_1\} \times \cdots \times \{X_d>x_d\}) dx_1\cdots dx_d.
\end{align}
In fact, for any ${\bf X}=(X_1, \cdots, X_d) \in \sX^d_+$, by Fubini's Theorem and the properties of $L_{\mu,{\bf X}}$,
\begin{align*}
   \Gamma_{\mu}({\bf X})
          & = \  \int_{\mathbb{R}^d_+} \int_{\mathbb{R}^d_+}
                                       1_{\{ {\bf y}<{\bf x} \}} d{\bf y} L_{\mu,{\bf X}} (d{\bf x})\\
          & = \  \int_{\mathbb{R}^d_+} \int_{\mathbb{R}^d_+}
                                       1_{\{ {\bf x}>{\bf y} \}} L_{\mu,{\bf X}} (d{\bf x}) d{\bf y}\\
          & = \  \int_0^{\infty}\cdots\int_0^{\infty} \mu(\{X_1>x_1\} \times \cdots \times \{X_d>x_d\}) dx_1\cdots dx_d.
\end{align*}
Moreover, from (\ref{3002}) it follows that for any ${\bf X} = (X_1, \cdots, X_d) \in  \sX^d_+$,
if $ \|X_i\| > 0$ for each $1 \leq i \leq d,$ then a sufficient condition ensuring $\Gamma_{\mu}({\bf X}) > 0$ is that
$\mu (A) > 0$ for any non-empty $A \in \sS.$

\vspace{0.3cm}

Note also that when $d=1$,
define a normalized and monotone set function $\mu$ on $\sF$ by $\mu(A):= g(P(A))$ with some distortion function $g$, then
$\Gamma_{\mu}$ reduces to
\begin{align*}
 \Gamma_{\mu}(X)= \int_0^{+\infty} g(P(X>x)) dx, \quad X \in \sX_+,
\end{align*}
which is just the distortion risk measure $\rho_g$ for the non-negative random variable $X$.
This is also the main reason why we call $\Gamma_{\mu}$ as in (\ref{3001}) the \textit{distortion} joint risk measure.

\vspace{0.3cm}

We would like to make one more comment about the definition of the distortion joint risk measure  $\Gamma_{\mu}$.
If one does not plan to discuss the axiomatic characterization of the distortion joint risk measure $\Gamma_{\mu}$,
then one can ignore the issues of the $d$-monotonicity and the continuity from below of $\mu,$
and thus can directly adopt formulation (\ref{3002}) as the definition of a distortion joint risk measure on $\sX^d_+.$
To be precise, let $Q$ be an arbitrarily given reference measure (scenario) on the measurable space $(\Omega, \sF).$ Then we can define
a normalized and monotone set function $\mu$ on $\sS$ by
\begin{align*}
\mu(A_1 \times \cdots \times A_d) := Q (A_1 \cdots A_d)), \quad  A_1 \times \cdots \times A_d \in \sS.
\end{align*}
Thus, the distortion joint risk measure $\Gamma_{\mu}$ given by (\ref{3002}) becomes
\begin{align*}
 \Gamma_Q({\bf X})
      := \int_0^{\infty}\cdots\int_0^{\infty} Q(X_1 > x_1, \cdots, X_d > x_d))  dx_1\cdots dx_d,  \quad  {\bf X}=(X_1, \cdots, X_d) \in \sX^d_+,
\end{align*}
which means that we can evaluate the joint risk of ${\bf X}$ under different reference measures $Q$ corresponding to plausible different
inter-dependence of $X_i's.$
Furthermore, assume that $g$ is a given distortion function.
Then we can define a normalized and monotone set function $\mu$ on $\sS$ by
\begin{align*}
\mu(A_1 \times \cdots \times A_d) := g( Q (A_1 \cdots A_d)), \quad  A_1 \times \cdots \times A_d \in \sS.
\end{align*}
Thus, the distortion joint risk measure $\Gamma_{\mu}$ given by (\ref{3002}) becomes
\begin{align*}
 \Gamma_{g}({\bf X})
      := \int_0^{\infty}\cdots\int_0^{\infty} g(Q(X_1 > x_1, \cdots, X_d > x_d))  dx_1\cdots dx_d,  \quad  {\bf X}=(X_1, \cdots, X_d) \in \sX^d_+,
\end{align*}
which is in accordance with the distortion risk measure for nonnegative multivariate risks proposed by Guillen et al. (2018, Section 5),
and also corresponds to the multivariate distortion risk measures proposed by R\"{u}schendorf (2006, Section 3) and  R\"{u}schendorf (2013, page 180).
Moreover, it is worth mentioning that, as pointed by Guillen et al. (2018, Introduction and Conclusions ),
a potential drawback may be the difficulty in interpreting their distortion risk measures $\Gamma_{g}$.
In the present paper, for the risk measures given by (\ref{3002}),
we have provided a plausible interpretation from the viewpoints of \textit{joint risk measures}.
In addition, we will also provide axiomatic characterization of the distortion joint risk measures given by (\ref{3002}),
see Theorems 3.2 and 3.3 below.
Taking the above considerations into account, the present study of distortion joint risk measures can be viewed as a meaningful complement to
the study of Guillen et al. (2018).

\vspace{0.3cm}

Now, we are ready to state the main results of this subsection.

\vspace{0.3cm}

\noindent{\bf Theorem 3.1} \ \ Let $\mu$ be a normalized, $d$-monotone and continuous from below set function on $\mathscr{S}.$
Then the distortion joint risk measure $\Gamma_{\mu}$ defined by ($\ref{3001}$) (or equivalently, as in ($\ref{3002}$))
is normalized, and satisfies the Axioms (A1)-(A5).

\vspace {0.3cm}

Next theorem provides an axiomatic characterization of the distortion joint risk measure $\Gamma_{\mu}$ defined as in ($\ref{3001}$)
(or equivalently, as in ($\ref{3002}$)).

\vspace {0.3cm}

\noindent{\bf Theorem 3.2} \ \ Suppose that a normalized joint risk measure $\Gamma: \sX^d_+ \rightarrow \mathbb{R}_+$
satisfies the Axioms (A1)-(A5). Then there exists a normalized, $d$-monotone and continuous from below set function $\mu$ on $\sS$
depending on $\Gamma,$
such that for any  ${\bf X}$ $ = $ $ (X_1, \cdots, X_d)$ $ \in $ $ \sX^d_+$,
\begin{align}\label{3003}
  \Gamma({\bf X})
    & = \int_{\mathbb{R}^d_+} x_1  \cdots  x_d L_{\mu,{\bf X}}(dx_1,\cdots,dx_d) \nonumber\\
    & = \int_0^{\infty}\cdots\int_0^{\infty} \mu(\{X_1>x_1\} \times \cdots \times \{X_d>x_d\}) dx_1\cdots dx_d.
\end{align}

\vspace{0.3cm}

\noindent{\bf Remark 3.2} \ \ From the proof of Theorem 3.2 below, we know that
\begin{align*}
  \mu ( A_1 \times \cdots \times A_d ) := \Gamma((1_{A_1},\cdots,1_{A_d})), \quad A_1 \times \cdots \times A_d \in \sS,
\end{align*}
which implies that $\mu$ is uniquely determined by $\Gamma.$ Furthermore, if we drop the normalization assumption on $\Gamma,$
then the above representation (\ref{3003}) becomes
\begin{align*}
  \Gamma({\bf X})
     = \Gamma({\bf 1}) \int_0^{\infty}\cdots\int_0^{\infty} \mu(\{X_1>x_1\} \times \cdots \times \{X_d>x_d\}) dx_1\cdots dx_d.
\end{align*}

\vspace{0.3cm}

\noindent{\bf Remark 3.3} \ \ When $d=1$,  the Axioms (A2)-(A5) become monotonicity, comonotone additivity and continuity from below, respectively.
Meanwhile, (\ref{3003}) reduces to
\begin{align*}
 \Gamma(X) = \int_{\mathbb{R}_+} x L_{\mu, X}(dx) = \int_0^{+\infty} \mu(X > x) dx, \quad X \in \sX_+,
\end{align*}
which is consistent with the univariate Greco's Representation Theorem; for instance, see Denneberg (1994, Theorem 13.2).
Notice also that in the univariate Greco's Representation Theorem, the monotonicity and
comonotone additivity are crucial for a functional to be representable as a Choquet integral, while weaker lower marginal and
upper marginal continuity are assumed instead of continuity from below. Thus, taking the above considerations into account,
Theorem 3.2 could also be mathematically viewed as a kind of multivariate extension of the univariate Greco's Representation Theorem.

\vspace{0.3cm}

Theorems 3.1 and 3.2 state that a normalized joint risk measure $\Gamma$ satisfies (A1)-(A5) if and only if it is
a distortion joint risk measure given as in (\ref{3001}) (or equivalently, as in (\ref{3002})).
Next, we will show that besides Axioms (A1)-(A5), if one further assumes Axiom (A6), then
for any continuous ${\bf X} \in \sX^d_+$, the distortion joint risk measure $\Gamma ({\bf X})$  will have a more explicit expression
in terms of certain copula and some distortion functions.
As a priori, we naturally further assume that the probability space $(\Omega, \sF, P)$ is large enough so that we can define any
continuous random vector on it.

\vspace{0.3cm}

\noindent{\bf Theorem 3.3}\ \ Suppose that a normalized joint risk measure $\Gamma: \sX^d_+ \rightarrow \mathbb{R}_+$
satisfies the Axioms (A1)-(A6). Then for any continuous ${\bf X} $ $ = $ $(X_1,$ $ \cdots,$ $ X_d)$ $ \in $ $ \sX^d_+ $ which has a unique copula $C$,
there exist left-continuous distortion functions $g_{_{1,C}}, \cdots, g_{_{d,C}}$ depending on $C$,
and a copula $C^*$ depending on $C$ which is uniquely determined on $\Pi_{i=1}^d \mbox{Ran}(g_{_{i,C}})$,
such that
\begin{align}\label{3004}
  \Gamma({\bf X}) =
     \int_0^{\infty}\cdots\int_0^{\infty} C^*\left(g_{_{1,C}} (P(X_1>t_1)), \cdots, g_{_{d,C}}(P(X_d>t_d))\right) dt_1 \cdots dt_d.
\end{align}

\vspace{0.3cm}

\noindent{\bf Remark 3.4} \ \ If we drop the normalization assumption on $\Gamma,$ then the above representation (\ref{3004}) becomes
\begin{align*}
  \Gamma({\bf X})
      = \Gamma({\bf 1})  \int_0^{\infty}\cdots\int_0^{\infty}
                                         C^*\left(g_{_{1,C}} (P(X_1>t_1)), \cdots, g_{_{d,C}}(P(X_d>t_d))\right) dt_1 \cdots dt_d.
\end{align*}
Moreover, when $d=1$, both the univariate copulas $C$ and $C^*$ are the identity function, that is, $C(x)=C^*(x) = x,$ $ x \in [0,1]$.
Hence, the distortion function $g_{_{1,C}}$ does not depend on the random variable $X \in \sX_+$ anymore,
and therefore (\ref{3004}) is in accordance with the classic distortion risk measures for non-negative random variables;
for instance, see Wang et al. (1997, Theorem 2). Taking above considerations into account, Theorem 3.3 can also be mathematically viewed as a
non-trivial multivariate generalization of univariate distortion risk measures.

\subsection{Vector-valued distortion joint risk measures}

In this subsection, we first establish a new class of vector-valued risk measures, which we refer to as
vector-valued distortion joint risk measures. Then we discuss their properties.

\vspace{0.3cm}

In general, a vector-valued joint risk measure is defined as any mapping $\tau : \sX^d_+ \longrightarrow \mathbb{R}^d_+$.
Recall that given a normalized joint risk measure $\Gamma :$ $\sX^d_+$ $ \longrightarrow $ $ \mathbb{R}_+$,
for any random vector ${\bf X} = (X_1, \cdots, X_d) \in \sX^d_+$, the quantity $\Gamma ({\bf X})$
represents the minimal amount of liquidity requirement or financial resource to cancel the joint risk of ${\bf X}.$
For each $1 \leq i \leq d,$ consider the random vector $X_i{\bf e}_i + \bar{{\bf e}}_i = (1, \cdots, 1, X_i, 1, \cdots, 1)$.
Notice that only the marginal $X_i$ is random in the random vector $X_i{\bf e}_i + \bar{{\bf e}}_i$, so it is natural to use the quantity
$ \Gamma (X_i{\bf e}_i + \bar{{\bf e}}_i)$ as sort of risk evaluation of $X_i$, if $X_i$ is considered as a marginal of ${\bf X}$ but not
as a stand-alone random variable.
For simplicity, write $\rho(X_i)$ for $ \Gamma (X_i{\bf e}_i + \bar{{\bf e}}_i), 1 \leq i \leq d.$ Hence, the quantity $\rho(X_i)$ could be
understood as the contribution of $X_i$ to the joint risk of $(1, \cdots, 1, X_i, 1, \cdots, 1)$.
Interestingly, it turns out that $\rho(X_i)$ is in accordance with some known vector-valued risk measures, respectively;
see Examples 4.1-4.4 below.
Motivated by above considerations, we naturally introduce the following definition of vector-valued distortion joint risk measures.

\vspace{0.3cm}

\noindent{\bf Definition 3.2}\ \ Let $\Gamma: \sX^d_+ \rightarrow \mathbb{R}_+$ be a normalized joint risk measure
satisfying the Axioms (A1)-(A5), and let $\mu$ be the normalized, $d$-monotone and continuous from below set function as in Theorem 3.2
such that (\ref{3003}) holds.
The vector-valued distortion joint risk measure $ H : \sX^d_+ \rightarrow \mathbb{R}^d_+ $ is defined by
\begin{align}\label{3005}
  H({\bf X})
     & :=  \left( \begin{array}{c}
                      \Gamma (X_1{\bf e}_1 + \bar{{\bf e}}_1) \\
                            \vdots \\
                      \Gamma (X_d{\bf e}_d + \bar{{\bf e}}_d)
                  \end{array}
           \right), \quad {\bf X} = (X_1, \cdots, X_d) \in \sX^d_+,
\end{align}
where for each $1 \leq i \leq d,$  $\Gamma (X_i{\bf e}_i + \bar{{\bf e}}_i)$ is given by
\begin{align}\label{221030add1}
 \Gamma (X_i{\bf e}_i + \bar{{\bf e}}_i)
       = \int_0^{+\infty} \mu(\Omega \times \cdots \times \Omega \times \{X_i> x_i\}\times \Omega \times \cdots \times \Omega)dx_i.
\end{align}

\vspace{0.3cm}

\noindent{\bf Remark 3.5} \ \ Besides the Axioms (A1)-(A5), if one further assumes Axiom (A6),
then for any continuous ${\bf X} = (X_1, \cdots, X_d) \in \sX^d_+$ having a copula $C$,
$H({\bf X})$ has the following expression:
\begin{align}\label{3006}
 H({\bf X})
   & = \left( \begin{array}{c}
                     \int_0^{+\infty} C^*(g_{_{1,C}}(P(X_1>x_1)), 1, \cdots, 1)dx_1 \\
                                   \vdots \\
                     \int_0^{+\infty} C^*(1, \cdots, 1, g_{_{d,C}}(P(X_d>x_d)))dx_d
              \end{array}
       \right) \nonumber\\  \nonumber\\
   & = \left(  \begin{array}{c}
                      \int_0^{+\infty} g_{_{1,C}}(P(X_1>x_1))dx_1 \\
                                    \vdots \\
                      \int_0^{+\infty} g_{_{d,C}}(P(X_d>x_d))dx_d
              \end{array}
      \right),
\end{align}
where the copula $C^*$ and the distortion functions $g_{_{1,C}}, \cdots, g_{_{d,C}}$ are as in Theorem 3.3.

\vspace{0.3cm}

The verification of (\ref{3006}) is quite similar to the arguments of the proof of Theorem 3.3 below. Hence,
we briefly demonstrate it here. Let us adopt all the notations used in the proof of Theorem 3.3.
More precisely, define $ U_i := F_{X_i}(X_i),$ $ 1\leq i \leq d,$ then $U_i$ is uniformly distributed on $(0,1)$.
Let $\gamma_{_C}$ be defined as in (\ref{221102add1}), the distortion function $g_{_{i,C}}$ be defined as in (\ref{221102add4}),
$1\leq i \leq d,$ and the copula $C^*$ be defined as in (\ref{221103add7}).
For each $1\leq i\leq d$,  by (\ref{221103add3}), (\ref{221103add4}), the definition of $\gamma_{_C}$
and the distribution invariance of $\Gamma,$  we have that for any $x_i \geq 0,$
\begin{align*}
    \mu & (\Omega \times \cdots \times \Omega \times \{X_i> x_i\} \times \Omega \times \cdots \times \Omega) \\
       & = \Gamma((1_{\{X_1>-1\}}, \cdots, 1_{\{X_{i-1}>-1\}}, 1_{\{X_i>x_i\}}, 1_{\{X_{i+1}>-1\}}, \cdots, 1_{\{X_d>-1\}})) \\
       & = \Gamma((1_{ \{U_1 > 0\} }, \cdots, 1_{ \{U_{i-1} > 0\} },
             1_{ \{U_i > F_{X_i}(x_i)\} }, 1_{ \{U_{i+1} > 0 \}}, \cdots, 1_{ \{U_d > 0\} })) \\
       & = \Gamma((1_{ \{U_1 > 1-1\} }, \cdots, 1_{ \{U_{i-1} > 1-1\} },
             1_{ \{U_i > 1-S_{X_i}(x_i)\} }, 1_{ \{U_{i+1} > 1-1 \}}, \cdots, 1_{ \{U_d > 1-1\} })) \\
       & = \gamma_{_C}(1, \cdots, 1, S_{X_i}(x_i), 1, \cdots, 1) \\
       & = C^*(g_{_{1,C}}(1), \cdots, g_{_{{i-1},C}}(1), g_{_{{i},C}}(S_{X_i}(x_i)), g_{_{{i+1},C}}(1), \cdots, g_{_{d,C}}(1)) \\
       & = C^*(1, \cdots, 1, g_{_{i,C}}(P(X_i>x_i)), 1, \cdots, 1) \\
       & = g_{_{i,C}}(P(X_i>x_i)),
\end{align*}
which, together with (\ref{221030add1}), yields  ($\ref{3006}$).

\vspace{0.3cm}

Next theorem discusses the properties of $H$ defined as in ($\ref{3005}$).

\vspace{0.3cm}

\noindent{\bf Theorem 3.4 }\ \ The vector-valued distortion joint risk measure $H$ satisfies
the following properties:
\begin{enumerate}
  \item[(1)] Positive homogeneity:  $H(c{\bf X}) = cH({\bf X})$ for any ${\bf X} \in \sX^d_+$ and any $c > 0$.

  \item[(2)] Translation invariance: $H({\bf X}+{\bf c}) = H({\bf X})+{\bf c}$ for any ${\bf X} \in \sX^d_+$ and ${\bf c} \in \mathbb{R}^d_+$.

  \item[(3)] Monotonicity: $H({\bf X}) \leq H({\bf Y})$ for any ${\bf X}, {\bf Y} \in \sX^d_+$ with ${\bf X} \leq {\bf Y}$.

  \item[(4)] Comonotone additivity: $H({\bf X}+{\bf Y}) = H({\bf X})+H({\bf Y})$ for any ${\bf X}$, ${\bf Y} \in \sX^d_+$, which are comonotone.
\end{enumerate}

\section{Comparisons}

In this section,
in order to reveal the relationship of the vector-valued distortion joint risk measure $H$ as in (\ref{3006}) with some known
vector-valued risk measures, we examine such known vector-valued risk measures as multivariate lower-orthant VaR of
Cousin and Di Bernardino (2013),  multivariate upper-orthant CTE of Cousin and Di Bernardino (2014),
multivariate tail conditional expectation of Landsman et al. (2016)
and multivariate tail distortion risk measures of Cai et al. (2017).

\vspace{0.3cm}

\noindent{\bf Example 4.1 (multivariate lower-orthant VaR)}\ \

\vspace{0.2cm}

Given any $C \in \sC$, which is absolutely continuous with respect to the Lebesgue measure on $[0,1]^d$, denote by $L_{C}$ the L-S measure induced
by $C$ on $([0,1]^d,\sB([0,1]^d))$. For $(t_1,\cdots,t_d)\in [0,1]^d$, $1\leq i\leq d$, define $U_i(t_1,\cdots,t_d):= t_i$, then we claim
that ${\bf{U}}_C:= (U_1,\cdots,U_d)$ is a random vector on $([0,1]^d,\sB([0,1]^d),L_C)$ with uniform marginal distribution functions and
joint distribution function $C$. In fact, for any $(t_1,\cdots,t_d) \in [0,1]^d$,
\begin{align*}
  L_C(U_1\leq t_1,\cdots,U_d\leq t_d)= \Delta_{(0,\cdots,0)}^{(t_1,\cdots,t_d)}C= C(t_1,\cdots,t_d),
\end{align*}
\begin{align*}
  L_C(U_i\leq t_i)= C(1,\cdots,1,t_i,1,\cdots,1)= t_i.
\end{align*}
Since ${\bf U}_C$ satisfies the regularity conditions of Cousin and Di Bernardino (2013),
the joint probability density function of ${\bf U}_C$ and $C({\bf U}_C)$ exists,
as well as the probability density function of $C({\bf U}_C)$,
which are denoted by $f_{({\bf U}_C,C({\bf U}_C))}(u_1,\cdots,u_d,\alpha)$ and $f_{C({\bf U}_C)}$, respectively.

\vspace{0.2cm}

For any ${\bf X} = (X_1, \cdots, X_d) \in \sX^d_+$,
denote by $F_{X_i}^{-1}$ the left-continuous inverse function of the distribution function $F_{X_i}$ of $X_i$, that is,
 $F_{X_i}^{-1}(u):= \inf \{ x\in \mathbb{R}_+: F_{X_i}(x)\geq u \}$, $ u \in (0,1),$ $1\leq i\leq d$.
For any $\alpha \in (0,1)$, if the fraction
\begin{align}\label{3007}
  \frac{\int_0^1\cdots\int_0^1 F_{X_1}^{-1}(u_1) \cdots  F_{X_d}^{-1}(u_d) f_{({\bf U}_C,C({\bf U}_C))}(u_1,\cdots,u_d,\alpha)du_1\cdots du_d}{f_{C({\bf U}_C)}(\alpha)}
\end{align}
makes sense, then denote it by
\begin{align*}
  E_{L_C}[F_{X_1}^{-1}(U_1) \cdots  F_{X_d}^{-1}(U_d)| C({\bf U}_C) = \alpha].
\end{align*}
Otherwise, define
\begin{align*}
  E_{L_C}[F_{X_1}^{-1}(U_1) \cdots F_{X_d}^{-1}(U_d)| C({\bf U}_C) = \alpha]:= 0,
\end{align*}
where $E_{L_C}(\cdot)$ stands for the expectation with respect to the L-S measure $L_C$.

\vspace{0.2cm}

Given any $\alpha \in (0,1)$, define a normalized joint risk measure $\Gamma_{\alpha}: \sX^d_+  \rightarrow \mathbb{R}_+$ by
\begin{align*}
 \Gamma_{\alpha}({\bf X}):= E_{L_C}[F_{X_1}^{-1}(U_1) \cdots F_{X_d}^{-1}(U_d)| C({\bf U}_C) = \alpha], \quad {\bf X} = (X_1, \cdots, X_d) \in \sX^d_+.
\end{align*}
Then, we can steadily verify that $\Gamma_{\alpha}$ satisfies the Axioms (A1)-(A6).

\vspace{0.2cm}

Given a continuous random vector ${\bf X} = (X_1, \cdots, X_d) \in \sX^d_+$ so that $C$ is the copula for it,
then by the proof of Theorem 3.3, we know that the corresponding copula $C^*$ in Theorem 3.3 is given by
\begin{align*}
  C^* (x_1, \cdots, x_d)
  = \frac{\int_{1-{g_{_{1,C}}^{-1}(x_1)}}^1 \cdots \int_{1-{g_{_{d,C}}^{-1}(x_d)}}^1 f_{({\bf U}_C, C({\bf U}_C))}(x_1,\cdots,x_d,\alpha)dx_1\cdots dx_d}{K'(\alpha)},
\end{align*}
$(x_1,\cdots,x_d)\in [0,1]^d$, and the corresponding distortion functions $g_{_{i,C}}$, $1\leq i\leq d$, are given by
\begin{align*}
  g_{_{i,C}}(1-u_i)= \frac{\int_{u_i}^1 f_{(U_i,C({\bf U}_C))}(x_i,\alpha)dx_i}{K'(\alpha)}, \quad u_i \in [0,1],
\end{align*}
where the Kendall function $K(\alpha):= L_C(C({\bf U}_C)\leq \alpha)$,
$K'(\alpha)$ is the first-order derivative of $K(\alpha)$,
and $f_{(U_i, C({\bf U}_C))}(x_i,\alpha)$ is the joint probability density function of the random vector $(U_i, C({\bf U}_C)), 1\leq i\leq d.$

\vspace{0.2cm}

Now, we claim that $H({\bf X})$ as in ($\ref{3006}$) with respect to $\Gamma_{\alpha}$ coincides with the multivariate lower-orthant VaR of
Cousin and Di Bernardino (2013). In fact, by Definition 3.2 and Remark 3.5, for each $1\leq i\leq d$,
\begin{align}\label{3008}
   \int_0^{+\infty} g_{_{i,C}}(P(X_i>x_i))dx_i
       = \frac{1}{K'(\alpha)}\int_0^{+\infty} \int_{F_{X_i}(x_i)}^1 f_{(U_i,C({\bf U}_C))}(u_i,\alpha)du_idx_i.
\end{align}

Next, we calculate the right-hand side of ($\ref{3008}$). For ${\bf x}= (x_1,\cdots,x_d)\in \mathbb{R}^d$,
\begin{align*}
  P & \left\{\omega : X_1(\omega) \leq x_1, \cdots, X_d(\omega) \leq x_d, P(X_1\leq X_1(\omega),
                 \cdots,X_d\leq X_d(\omega))\leq \alpha \right\}\\
    & = P \left\{ \omega : F_{X_1}(X_1)(\omega)\leq F_{X_1}(x_1), \cdots, F_{X_d}(X_d)(\omega)\leq F_{X_d}(x_d),\right.\\
     & \left. \qquad \qquad \qquad \qquad \qquad C(P(X_1\leq X_1(\omega)),\cdots,P(X_d\leq X_d(\omega)))\leq \alpha \right\}\\
    & = P \left\{ \omega : \widetilde{U_1}(\omega)\leq u_1, \cdots, \widetilde{U_d}(\omega)\leq u_d,
              C(\widetilde{U_1}(\omega),\cdots,\widetilde{U_d}(\omega))\leq \alpha \right\},
\end{align*}
where $\widetilde{U_i} := F_{X_i}(X_i)$ and $u_i :=F_{X_i}(x_i)$, $1\leq i\leq d$.
Denote $\widetilde{{\bf U}} := (\widetilde{U_1},\cdots,\widetilde{U_d})$.
Then by Lemma 2.1(1), we know that $C$ is a copula for $\widetilde{{\bf U}}$.
Notice the fact that $\widetilde{{U_i}}$ and $U_i$ have the same distribution
function, $1\leq i\leq d$, hence $\widetilde{{\bf U}}$ has the same joint distribution function as that of ${\bf U}_C$. Thus,
\begin{align*}
P & \left\{ \omega : \widetilde{U_1}(\omega)\leq u_1, \cdots, \widetilde{U_d}(\omega)\leq u_d,
      C(\widetilde{U_1}(\omega),\cdots,\widetilde{U_d}(\omega))\leq \alpha \right\} \\
   & = P\circ{\widetilde{{\bf U}}^{-1}}\left\{ {\bf t}= (t_1,\cdots,t_d)\in [0,1]^d: 0\leq t_1\leq u_1, \cdots, 0\leq t_d\leq u_d, C(t_1,\cdots,t_d)
       \leq \alpha \right\}\\
   & = L_{C} \left\{ {\bf t}= (t_1,\cdots,t_d)\in [0,1]^d: t_1\leq u_1, \cdots, t_d\leq u_d, C(t_1,\cdots,t_d)\leq \alpha \right\}\\
   & = L_{C} \left\{ {\bf t} \in [0,1]^d: {U_1}({\bf t})\leq u_1, \cdots, {U_d}({\bf t})\leq u_d, C({U_1},\cdots,{U_d})({\bf t})\leq \alpha \right\}.
\end{align*}
Therefore,
\begin{align*}
   P & \left\{ \omega : X_1(\omega) \leq x_1, \cdots, X_d(\omega) \leq x_d, P(X_1\leq X_1(\omega),\cdots,X_d\leq X_d(\omega))\leq \alpha \right\}\\
   & = L_{C} \left\{ {\bf t} \in [0,1]^d: {U_1}({\bf t})\leq u_1, \cdots, {U_d}({\bf t})\leq u_d, C({U_1},\cdots,{U_d})({\bf t})\leq \alpha \right\},
\end{align*}
which implies that
\begin{align*}
  F_{(X_i,F_{\bf X}({\bf X}))}(x_i,\alpha)= F_{(U_i,C({\bf U}_C))}(u_i,\alpha).
\end{align*}
Therefore,
\begin{align}\label{add3015}
  & f_{(X_i,F_{\bf X}({\bf X}))}(x_i,\alpha)= f_{(U_i,C({\bf U}_C))}(u_i,\alpha)f_{X_i}(x_i),
\end{align}
\begin{align}\label{hyjadd3016}
  & P(F_{\bf X}({\bf X})\leq \alpha)= K(\alpha),
\end{align}
and when $x_i\leq F_{X_i}^{-1}(\alpha)$, $f_{(X_i,F_{\bf X}({\bf X}))}(x_i,\alpha)=0$.
Consequently, it follows from ($\ref{3008}$), ($\ref{add3015}$) and ($\ref{hyjadd3016}$) that
\begin{align*}
  \int_0^{+\infty} g_{_{i,C}}(P(X_i>x_i))dx_i
          & = \frac{1}{f_{F_{\bf X}({\bf X})}(\alpha)}\int_0^{+\infty} \int_{F_{X_i}(x_i)}^1 f_{(U_i,C({\bf U}_C))}(u_i,\alpha)du_idx_i\\
          & = \frac{1}{f_{F_{\bf X}({\bf X})}(\alpha)}\int_0^{+\infty} \int_{x_i}^{+\infty} f_{(X_i,F_{\bf X}({\bf X}))}(y_i,\alpha)dy_idx_i\\
          & = \frac{1}{f_{F_{\bf X}({\bf X})}(\alpha)}\int_0^{+\infty} \int_0^{y_i} f_{(X_i,F_{\bf X}({\bf X}))}(y_i,\alpha)dx_idy_i\\
          & = \frac{1}{f_{F_{\bf X}({\bf X})}(\alpha)}\int_0^{+\infty} x_i\cdot f_{(X_i,F_{\bf X}({\bf X}))}(x_i,\alpha)dx_i\\
          & = \frac{1}{f_{F_{\bf X}({\bf X})}(\alpha)}\int_{F_{X_i}^{-1}(\alpha)}^{+\infty} x_i\cdot f_{(X_i,F_{\bf X}({\bf X}))}(x_i,\alpha)dx_i,
\end{align*}
which coincides with the multivariate lower-orthant VaR of Cousin and Di Bernardino (2013).

\vspace{0.3cm}

\noindent{\bf Example 4.2 (multivariate upper-orthant CTE)}\ \

\vspace{0.2cm}

Suppose that a copula $C \in \sC$ is arbitrarily given, which is absolutely continuous with respect to the Lebesgue measure on $[0,1]^d$.
Given an $\alpha \in (0,1)$, define a normalized joint risk measure $\Gamma_{\alpha}: \sX^d_+  \rightarrow \mathbb{R}_+$ by
\begin{align*}
  \Gamma_{\alpha}({\bf X}):=
             E_{L_C}[F_{X_1}^{-1}(U_1) \cdots  F_{X_d}^{-1}(U_d)| S_{{\bf U}_C}({\bf U}_C)\leq 1-\alpha], \quad {\bf X} = (X_1, \cdots, X_d) \in \sX^d_+,
\end{align*}
where ${\bf U}_C := (U_1,\cdots,U_d)$, $L_C$ and the inverse functions $F_{X_1}^{-1}, 1\leq i\leq d,$ are as in Example 4.1.
Then, we can steadily verify that $\Gamma_{\alpha}$ satisfies the Axioms (A1)-(A6).

\vspace{0.2cm}

Given any continuous ${\bf X} \in \sX^d_+$ so that $C$ is a copula for it, then by the proof of Theorem 3.3, we know that the corresponding
copula $C^*$ in Theorem 3.3 is given by
\begin{align*}
 C^* (x_1, \cdots, x_d)
     & = \frac{1}{L_C(S_{{\bf U}_C}({\bf U}_C)\leq 1-\alpha)}\\
     & \quad
            \cdot \int_{1-g_{_{1,C}}^{-1}(x_1)}^1\cdots \int_{1-g_{_{d,C}}^{-1}(x_d)}^1 \int_0^{1-\alpha}
                                  f_{({\bf U}_C, S_{{\bf U}_C}({\bf U}_C))}(u_1,\cdots,u_d,y) dydu_1\cdots du_d,
\end{align*}
and the corresponding distortion functions $g_{_{i,C}}$, $1\leq i\leq d$, are given by
\begin{align*}
  g_{_{i,C}}(u_i) = \frac{1}{L_C(S_{{\bf U}_C}({\bf U}_C)\leq 1-\alpha)}
                               \cdot \int_{1-u_i}^1 \int_0^{1-\alpha} f_{(U_i,S_{{\bf U}_C}({\bf U}_C))}(x_i,y)dydx_i,
\end{align*}
where $f_{({\bf U}_C, S_{{\bf U}_C}({\bf U}_C))}$ is the joint probability density function of ${\bf U}_C$ and $S_{{\bf U}_C}({\bf U}_C)$, and
$f_{(U_i, S_{{\bf U}_C}({\bf U}_C)}$ is the joint probability density function of $U_i$ and $S_{{\bf U}_C}({\bf U}_C)$, $1\leq i\leq d$.

\vspace{0.2cm}

By Definition 3.2 and Remark 3.5, an elementary calculation shows that the $i$th component of $H({\bf X})$ as in ($\ref{3006}$) is given by
\begin{align*}
  \int_0^{\infty} g_{_{i,C}}(P(X_i>x_i))dx_i= \int_0^{+\infty} \int_{F_{X_i}(x_i)}^1 \int_0^{1-\alpha} f_{U_i,S_{{\bf U}_C}({\bf U}_C)}(u_i,y)dydu_idx_i,
\end{align*}
which, by a similar arguments as in Example 4.1, can be shown to coincide with the multivariate upper-orthant CTE of Cousin and Di Bernardino (2014).

\vspace{0.3cm}

\noindent{\bf Example 4.3 (multivariate tail conditional expectation, MTCE)}\ \

\vspace{0.2cm}

Suppose that a copula $C \in \sC$ is arbitrarily given, which is absolutely continuous with respect to the Lebesgue measure on $[0,1]^d$.
Given a $q \in (0,1)$, define a normalized joint risk measure $\Gamma_q: \sX^d_+ \rightarrow \mathbb{R}_+$ by
\begin{align*}
 \Gamma_q({\bf X}):=
  E_{L_C}\left[F_{X_1}^{-1}(U_1) \cdots  F_{X_d}^{-1}(U_d)| U_1 > {\rm VaR}_q (U_1), \cdots, U_d > {\rm VaR}_q (U_d)\right],
\end{align*}
${\bf X} = (X_1, \cdots, X_d) \in \sX^d_+,$
where ${\bf U}_C := (U_1,\cdots,U_d)$, $L_C$ and the left-continuous inverse functions $F_{X_1}^{-1}, 1\leq i\leq d,$ are as in Example 4.1,
${\rm VaR}_q (U_i) := F_{U_i}^{-1}(q), 1\leq i\leq d.$
Then, we can steadily verify that $\Gamma_{q}$ satisfies the Axioms (A1)-(A6).

\vspace{0.2cm}

Given any continuous ${\bf X} \in \sX^d_+$ so that $C$ is a copula for it, then by the proof of Theorem 3.3, we know that the corresponding
copula $C^*$ in Theorem 3.3 is given by
\begin{align*}
   C^* (u_1, \cdots, u_d) = \frac{1}{p} \widehat{C}(g_{_{1,C}}^{-1}(u_1)\wedge \alpha, \cdots, g_{_{d,C}}^{-1}(u_d)\wedge \alpha),
\end{align*}
and the corresponding distortion functions $g_{_{i,C}}$, $1\leq i\leq d$, are given by
\begin{align*}
  g_{_{i,C}}(u_i)=
  \begin{cases}
  \frac{\widehat{C}(\alpha, \cdots, \alpha, u_i, \alpha, \cdots, \alpha)}{\widehat{C}(\alpha, \cdots, \alpha)}, & 0 \leq u_i \leq \alpha, \\
  1, & \alpha < u_i \leq 1,
  \end{cases}
\end{align*}
where $\widehat{C}$ is the survival copula associated with copula $C$, $\alpha := 1-q $ and
\begin{align*}
  p:&=\widehat{C}(P(X_1 > {\rm VaR}_q (X_1)), \cdots, P(X_d > {\rm VaR}_q (X_d)))\\
  &=\widehat{C}(1-q, \cdots, 1-q)\\
  &=\widehat{C}(\alpha, \cdots, \alpha).
\end{align*}

\vspace{0.2cm}

By Definition 3.2 and Remark 3.5, an elementary calculation shows that the $i$th component of $H({\bf X})$ as in ($\ref{3006}$) is given by
\begin{align*}
  \int_0^{\infty} g_{_{i,C}}(P(X_i > x_i))dx_i= \frac{1}{\widehat{C}(\alpha,\cdots,\alpha)}
              \int_0^{+\infty} \widehat{C}(\alpha,\cdots,\alpha,P(X_i > x_i),\alpha,\cdots,\alpha)dx_i,
\end{align*}
which, by a similar arguments as in Example 4.1, can be shown to coincide with the MTCE of Landsman et al. (2016).

\vspace{0.3cm}

\noindent{\bf Example 4.4 (multivariate tail distortion risk measure, MTDRM)}\ \

\vspace{0.2cm}

Suppose that a copula $C \in \sC$ is arbitrarily given, and that $\widehat{C}$ is the survival copula associated with the copula $C$.
Given $d$ distortion functions $g_1,\cdots,g_d$, define a normalized joint risk measure
$\Gamma :$  $\sX^d_+ \rightarrow \mathbb{R}_+$ by
\begin{align*}
  \Gamma &({\bf X}) :=
     \int_{[0,+\infty)^d} \frac{\widehat{C}(g_1( P(\Omega_{\bf X} \cap \{ X_1>x_1 \})),\cdots,g_d( P(\Omega_{\bf X} \cap \{ X_d>x_d \})))
         dx_1\cdots dx_d}{P(\Omega_{\bf X})},
\end{align*}
${\bf X} = (X_1, \cdots, X_d) \in \sX^d_+,$
where $\Omega_{\bf X}$ is a tail region defined as in Cai et al. (2017).
Assume that if ${\bf X} = (X_1, \cdots, X_d)$ and ${\bf Y} = (Y_1, \cdots, Y_d) \in \sX^d_+$ have the same copula,
then for any $(u_1,\cdots,u_d) \in [0,1]^d$,
\begin{align*}
 P(\Omega_{\bf X}\cap\{ F_{X_i}(X_i)>u_i, i=1,\cdots,d \}) = P(\Omega_{\bf Y}\cap \{ F_{Y_i}(Y_i)>u_i, i=1,\cdots,d \}).
\end{align*}
Note that this assumption holds when $ \Omega_{\bf X}$ and $ \Omega_{\bf Y}$ are specified to the sample space $\Omega$.
Under the above assumption, we can steadily verify that  $\Gamma$ satisfies the Axioms (A1)-(A6).

\vspace{0.2cm}

For any continuous ${\bf X} = (X_1, \cdots, X_d) \in \sX^d_+$ so that $C$ is a copula for it, then by the proof of Theorem 3.3,
we know that the corresponding copula $C^*$ in Theorem 3.3 is given by
\begin{align*}
     & C^* (x_1, \cdots, x_d)\\
     & = \  \frac{1}{P({\Omega}_{\bf X})} \\
     &\quad \cdot \widehat{C}(g_1( P({\Omega}_{\bf X} \cap \{ F_{X_1}(X_1)>1-{g_{_{1,C}}^{-1}(x_1)} \}),\cdots,g_d( P({\Omega}_{\bf X} \cap \{ F_{X_d}(X_d)>1-{g_{_{d,C}}^{-1}(x_d)} \})),
\end{align*}
and the corresponding distortion functions $g_{_{i,C}}$, $1\leq i\leq d$, are given by
\begin{equation*}
  g_{_{i,C}}(1-u_i)= \frac{1}{P({\Omega}_{\bf X})} g_i( P({\Omega}_{\bf X} \cap \{ F_{X_i}(X_i)>u_i \})).
\end{equation*}
By Definition 3.2 and Remark 3.5, an elementary calculation shows that the  vector-valued distortion joint risk measure
$H({\bf X})$ as in ($\ref{3006}$) is given by
\begin{align*}
H({\bf X})= (H_{g_1,\Omega_{\bf X}}(X_1),\cdots,H_{g_d,\Omega_{\bf X}}(X_d))^t,
\end{align*}
where the superscript $t$ stands for transpose of a vector, and for each $1\leq i\leq d,$
\begin{align*}
H_{g_i,\Omega_{\bf X}}(X_i) := \frac{1}{P(\Omega_{\bf X})} \int_{0}^{+\infty} g_i ( P(\Omega_{\bf X}\cap \{ X_i> x_i \}))dx_i,
\end{align*}
which coincides with the MTDRM of Cai et al. (2017).

\section{Concluding remarks}

In order to measure the \textit{joint risk} of multiple insurance risks (i.e. of non-negative random vectors), we suggest and characterize
distortion joint risk measures for non-negative random vectors. Inspired by both the theory of expected utility and Cobb-Dauglas utility function,
we establish a (scalar) distortion joint risk measure to quantify the \textit{joint risk} of non-negative random vectors.
To the best of our knowledge, distortion joint risk measures are new type of risk measures.
By proposing a set of new axioms, we also provide axiomatic characterization of distortion joint risk measures.
The most novel axiom is the component-wise positive homogeneity.
Furthermore,  we also introduce and study a new class of vector-valued distortion joint risk measures.
Comparison study shows that some known vector-valued multivariate risk measures are of the forms of vector-valued distortion joint risk measures,
respectively. This paper mainly gives some theoretical results about the evaluation of joint risk under dependence uncertainty,
and it is anticipated to be helpful to measure the joint risk of multivariate risks under model uncertainty.

\vspace{0.2cm}

One of the employed axioms is the comonotone additivity. This axiom is equivalent to using a model
where the random vectors are assumed to be comonotone (i.e. non-diversified), and it represents the worst-case situation.
A natural and interesting issue could be to take diversification advantage into account.
In other words, one could replace the comonotone additivity with the subadditivity,
and it would be interesting to see relevant study to be worked out somehow in the future.

\section*{Acknowledgements}

The authors are very grateful to Professor Hailiang Yang for his very helpful discussions and comments on an earlier version of this manuscript.

\section*{Appendix}
\setcounter{equation}{0}
\setcounter{subsection}{0}
\renewcommand{\theequation}{A.\arabic{equation}}
\renewcommand{\thesubsection}{A.\arabic{subsection}}

\subsection{\textbf{Proofs}}

In this section, we provide proofs of all the main results of this paper.

\vspace{0.2cm}

\noindent{\bf Proof of Lemma 2.1}

\vspace{0.2cm}

Let a random vector ${\bf X} = (X_1, \cdots, X_d) \in \sX^d_+$ have a copula $C.$

\vspace{0.2cm}

(1) For $1\leq i\leq d$, define the right-continuous inverse function $g_i^{-1}$ of $g_i$ as
\begin{align*}
    g_i^{-1}(u_i):= \inf \{ x\in\mathbb{R}: g_i(x)>u_i \}
\end{align*}
and $\inf{\emptyset}:= +\infty$. Then for any $x_i \in \mbox{Ran}(g_i),$
\begin{align*}
  \{g_i(X_i)\leq x_i\}= \{ X_i \leq g_i^{-1}(x_i)\}.
\end{align*}
Thus,
\begin{align*}
  P&(g_1(X_1)\leq x_1,\cdots,g_d(X_d)\leq x_d)\\
  &= P(X_1 \leq g_1^{-1}(x_1),\cdots, X_d \leq g_d^{-1}(x_d))\\
  &= C(P(X_1 \leq g_1^{-1}(x_1)),\cdots,P(X_d \leq g_d^{-1}(x_d)))\\
  &= C(P(g_1(X_1)\leq x_1),\cdots,P(g_d(X_d)\leq x_d)).
\end{align*}

(2) Given an $(a_1, \cdots, a_d) \in \mathbb{R}^d,$ for each $1\leq i\leq d$, if $0\leq x_i< 1$, then
\begin{align*}
  \{1_{\{X_i>a_i\}}\leq x_i \} = \{ X_i\leq a_i \}.
\end{align*}
If $x_i<0$, then
\begin{align*}
  \{ 1_{\{X_i>a_i\}}\leq x_i \} = \emptyset = \{ X_i\leq -\infty \}.
\end{align*}
If $x_i \geq 1$, then
\begin{align*}
  \{ 1_{\{X_i>a_i\}}\leq x_i \} = \Omega = \{ X_i\leq +\infty \}.
\end{align*}
For each $1\leq i\leq d,$ define a function $f: \mathbb{R} \rightarrow \mathbb{R}\cup\{+\infty\} \cup\{-\infty\}$ by
\begin{align*}
 f_i(x):=
    \begin{cases}
       a_i,     & \mbox{if}\ 0\leq x < 1,\\
       -\infty, & \mbox{if}\ x < 0,\\
       +\infty, & \mbox{if}\ x \geq 1.\\
   \end{cases}
\end{align*}
Hence, for any $(x_1, \cdots, x_d) \in \mathbb{R}^d,$
\begin{align*}
 P&(1_{\{X_1>a_1\}}\leq x_1,\cdots,1_{\{X_d>a_d\}}\leq x_d)\\
  & = P(X_1 \leq f_1(x_1),\cdots,X_d \leq f_d(x_d))\\
  & = C(P(X_1 \leq f_1(x_1)),\cdots,P(X_d \leq f_d(x_d)))\\
  & = C(P(1_{\{X_1>a_1\}}\leq x_1),\cdots, P(1_{\{X_d>a_d\}}\leq x_d)).
\end{align*}
Lemma 2.1 is proved.

\vspace{0.5cm}

\noindent{\bf Proof of Lemma 2.2}

\vspace{0.2cm}

(1) Note that for any real numbers $a<b$, the function
\begin{align*}
 h_{b,a}(x):= x\wedge b- x\wedge a=
   \begin{cases}
       0,   & \mbox{if}\ x\leq a,\\
       x-a, & \mbox{if}\ a<x \leq b,\\
       b-a, & \mbox{if}\ x>b,
   \end{cases}
\end{align*}
is continuous and non-decreasing. Hence, by Remark 2.1, we know that for each $1\leq i\leq d$,
\begin{align*}
X_i^k := \sum\limits_{j_i=1}^{k_i} \left( X_i \wedge \frac{j_i+1}{2^{n_i}} - X_i \wedge \frac{j_i}{2^{n_i}}\right)
\end{align*}
and
\begin{align*}
X_i^l := \sum\limits_{j_i=1}^{l_i} \left( X_i \wedge \frac{j_i+1}{2^{n_i}} - X_i \wedge \frac{j_i}{2^{n_i}}\right).
\end{align*}
are comonotone. Consequently, the desired assertion is shown.

\vspace{0.2cm}

(2) By the same arguments of proving (1), we can show the desired assertion.

\vspace{0.2cm}

(3) By Lemma 2.1, we know that the desired assertions hold. Lemma 2.2 is proved.

\vspace{0.3cm}

\noindent{\bf Proof of Remark 3.1}

\vspace{0.2cm}

To show Remark 3.1, we first claim that from Axiom (A3)
it follows that $\tau({\bf X}) = 0$, if there is a zero-valued component of ${\bf X}$.
Indeed, without any loss of generality, we assume that $X_d = 0$, that is, ${\bf X} = (X_1, \cdots, X_{d-1}, 0).$
Note that any two random vectors of all random vectors involved later are comonotone.
Consider the random vector $4{\bf X}$. On the one hand,  by Axiom (A3) we know that
\begin{align}\label{add092501}
 \tau(4{\bf X}) = \tau(2{\bf X} + 2{\bf X})
                = 2^d \tau(2{\bf X}) = 2^d \tau({\bf X} + {\bf X})
                = 2 ^{2d} \tau({\bf X}).
\end{align}
On the other hand, by Axiom (A3) again, we have that
\begin{align}\label{add092502}
\tau (4{\bf X})
     & = \tau( (2X_1, \cdots, 2X_{d-2}, X_{d-1}, 0) + (2X_1, \cdots, 2X_{d-2}, 3X_{d-1}, 0)) \nonumber \\
     & = 2^{d-1} \tau( (2X_1, \cdots, 2X_{d-2}, X_{d-1}, 0)) + 2^{d-1} \tau( (2X_1, \cdots, 2X_{d-2}, 3X_{d-1}, 0)).
\end{align}
Note that by Axiom (A3),
\begin{align}\label{add092503}
 \tau( & (2X_1, \cdots, 2X_{d-2}, X_{d-1}, 0)) \nonumber \\
       & = \tau( (X_1, \cdots, X_{d-2}, X_{d-1}, 0) + (X_1, \cdots, X_{d-2}, 0, 0)) \nonumber \\
       & =2^{d-1} \tau( (X_1, \cdots, X_{d-2}, X_{d-1}, 0)) + 2^{d-1} \tau( (X_1, \cdots, X_{d-2}, 0, 0)),
\end{align}
\begin{align}\label{add092504}
 \tau( & (2X_1, \cdots, 2X_{d-2}, 3X_{d-1}, 0)) \nonumber \\
       & = \tau( (X_1, \cdots, X_{d-2}, X_{d-1}, 0) + (X_1, \cdots, X_{d-2}, 2X_{d-1}, 0)) \nonumber \\
       & = 2^{d-1} \tau( (X_1, \cdots, X_{d-2}, X_{d-1}, 0)) + 2^{d-1} \tau( (X_1, \cdots, X_{d-2}, 2X_{d-1}, 0))
\end{align}
and
\begin{align}\label{add092505}
 \tau(  (X_1, \cdots, X_{d-2}, 2X_{d-1}, 0))
       & = \tau( (X_1, \cdots, X_{d-2}, X_{d-1}, 0) + (0, \cdots, 0, X_{d-1}, 0)) \nonumber \\
       & \geq 2^2 \tau( (X_1, \cdots, X_{d-2}, X_{d-1}, 0)),
\end{align}
where the non-negativity of $\tau$ is used. From (\ref{add092502})-(\ref{add092505}), it follows that
\begin{align*}
 \tau & (4{\bf X}) \geq (2^{2d} + 2^{2d-1}) \tau({\bf X}),
\end{align*}
which, together with (\ref{add092501}), yields that $\tau({\bf X}) = 0.$

\vspace {0.2cm}

Second, we further claim that $\tau ((X_1, \cdots, X_{d-1}, c_d X_d)) = c_d \tau({\bf X})$
for any positive integer $c_d \geq 1$ and any ${\bf X} = (X_1, \cdots,X_d) \in \sX^d_+$.
In fact, by Axiom (A3) and previous claim, we know that
\begin{align*}
 \tau (  (X_1, \cdots, X_{d-1}, c_d X_d))
        & =  \tau ((X_1, \cdots, X_{d-1}, X_d) + (0, \cdots, 0, (c_d-1)X_d)) \\
        & = \tau ((X_1, \cdots, X_{d-1}, X_d)) + \tau ((X_1, \cdots, X_{d-1}, (c_d-1)X_d))\\
        & = c_d \tau ( {\bf X}).
\end{align*}
Similarly, we can show that Axiom (A1) holds for any positive integers $c_1, \cdots, c_d$.
Finally, under the help of monotonicity (A2), by an approximation approach similar to the arguments as in
Lemma 4.83 of F\"{o}llmer and Schied (2016) or Exercise 11.1 of Denneberg (1994), we can steadily show that Axiom (A1) holds
for positive rational and irrational numbers $c_1, \cdots, c_d,$ consecutively. Remark 3.1 is proved.

\vspace {0.3cm}

\noindent{\bf Proof of Theorem 3.1}

\vspace{0.2cm}

By ($\ref{3002}$) and the monotonicity of $\mu$, the monotonicity of $\Gamma_{\mu}$ is clear.
Since $\mu$ is continuous from below, by (\ref{3002}) we can easily know that $\Gamma_{\mu}$ is continuous from below.

\vspace{0.2cm}

We now turn to show the comonotone additivity. Given comonotone ${\bf X} = (X_1,\cdots,X_d)$ and $ {\bf Y} = (Y_1,\cdots,Y_d)$ $ \in $ $ \sX^d_+$,
by Remark 2.1, there exist continuous and non-decreasing functions $u_1,\cdots,u_d$, $v_1,\cdots,v_d$ on $\mathbb{R},$ such that
$u_i(u)+v_i(u)=u$, $u\in \mathbb{R}$ and $X_i  = u_i(X_i+Y_i)$, $ Y_i  = v_i(X_i+Y_i), 1\leq i\leq d$.
By ($\ref{3002}$), we have that
\begin{align*}
 \Gamma_{\mu}  ({\bf X}+{\bf Y})
      & :=\int_{\left[0,+\infty\right)^d} x_1  \cdots  x_d L_{\mu,{\bf X+Y}}(dx_1,\cdots,dx_d)\\
      & \ = \int_0^{+\infty}\cdots\int_0^{+\infty} \mu \left( \{ X_1+Y_1 > x_1 \} \times \cdots \times \{ X_d+Y_d > x_d \} \right) dx_1 \cdots dx_d.
\end{align*}
To calculate above integral. Given any $x_1, \cdots, x_d \geq 0$ and any $ {\bf W} = (W_1, \cdots, W_d) \in \sX^d_+,$ for each $1\leq i\leq  d,$
consider the monotone set function $\nu_i : \sF \rightarrow [0, 1]$ defined by
\begin{align*}
 \nu_i(A) := \mu \left( \{ W_1 > x_1 \} \times \cdots \times \{ W_{i-1} > x_{i-1} \} \times A \times \{ W_{i+1} > x_{i+1} \}
                          \times \cdots \times \{ W_d > x_d \} \right).
\end{align*}
Then for any non-negative random variable $V \in \sX_+,$ the Choquet integral $\int V d\nu_i$ of $V$ with respect to $\nu_i$ is
\begin{align*}
 \int V d\nu_i
   & := \int_0^{+\infty} \mu ( \{ W_1 > x_1 \} \times \cdots \times \{ W_{i-1} > x_{i-1} \} \times \{ V > t_i \} \times \{ W_{i+1} > x_{i+1} \} \\
   & \qquad \qquad \quad \quad  \times \cdots \times \{ W_d > x_d \} ) dt_i.
\end{align*}
By the comonotone additivity of Choquet integral (for instance, see Proposition 4.86 and Lemma 4.90 of F\"{o}llmer and Schied (2016),
or Proposition 5.1(vi) of Denneberg (1994)), we know that
\begin{align*}
  \int (X_i + Y_i) d\nu_i = \int X_i d\nu_i + \int Y_i d\nu_i,
\end{align*}
which exactly says that
\begin{align}\label{1010add1}
   & \int_0^{+\infty} \mu ( \{ W_1 > x_1 \} \times \cdots \times \{ W_{i-1} > x_{i-1} \} \times \{ X_i + Y_i > t_i \} \times \{ W_{i+1} > x_{i+1} \}
                                                                                                                                                \nonumber \\
   & \qquad \quad \quad  \times \cdots \times \{ W_d > x_d \} ) dt_i \nonumber \\
   & \quad =  \int_0^{+\infty} \mu ( \{ W_1 > x_1 \} \times \cdots \times \{ W_{i-1} > x_{i-1} \} \times \{ X_i > t_i \}
                                 \times \{ W_{i+1} > x_{i+1} \} \nonumber \\
   & \qquad \qquad \qquad \quad \times \cdots \times \{ W_d > x_d \} ) dt_i \nonumber \\
   & \qquad + \int_0^{+\infty} \mu ( \{ W_1 > x_1 \} \times \cdots \times \{ W_{i-1} > x_{i-1} \} \times \{ Y_i > t_i \}
                                 \times \{ W_{i+1} > x_{i+1} \} \nonumber \\
   & \qquad \qquad \qquad \quad \times \cdots \times \{ W_d > x_d \} ) dt_i.
\end{align}

Applying consecutively (\ref{1010add1}) to $ i = 1, \cdots, d$ yields that
\begin{align*}
   &  \int_0^{+\infty}\cdots\int_0^{+\infty}\mu \left( \{ X_1+Y_1 > x_1 \} \times \cdots \times \{ X_d+Y_d > x_d \} \right) dx_1 dx_2 \cdots dx_d \\
   & = \int_0^{+\infty}\cdots\int_0^{+\infty} \mu \left(
                            \{ X_1 > x_1 \} \times \{ X_2+Y_2 > x_2 \} \times \cdots \times \{ X_d+Y_d > x_d \} \right) dx_1dx_2 \cdots dx_d \\
   &\quad + \int_0^{+\infty}\cdots\int_0^{+\infty} \mu \left(
                            \{ Y_1 > y_1 \} \times \{ X_2+Y_2 > x_2 \} \times \cdots \times \{ X_d+Y_d > x_d \} \right) dy_1dx_2 \cdots dx_d \\
   & =  \underset{{\bf Z}}{\sum} \int_0^{+\infty}\cdots\int_0^{+\infty} \mu \left(
                            \{ Z_1 > x_1 \} \times \cdots \times \{ Z_d > x_d \} \right) dx_1 \cdots dx_d \\
   & =  \underset{{\bf Z}}{\sum} \Gamma_{\mu}((Z_1, \cdots, Z_d)),
\end{align*}
where the summation $\underset{{\bf Z}}{\sum}$ is summed over the $2^d$ random vectors ${\bf Z} = (Z_1, \cdots, Z_d)$ with $Z_i$ being
either $X_i$ or $Y_i$ for each $1\leq i\leq d.$
The comonotone additivity of $ \Gamma_{\mu}$ is shown.

\vspace{0.2cm}

Finally, we show the  $d$-monotonicity of $\Gamma_{\mu}$. Given ${\bf X}=(X_1,\cdots,X_d)$, ${\bf Y}=(Y_1,\cdots,Y_d) \in \sX^d_+$
with ${\bf X} \leq {\bf Y}$, denote ${\bf Z} = (Z_1,\cdots,Z_d)$ with $Z_i$ being either $X_i$ or $Y_i$ for each $1 \leq i \leq d$, and
denote by $\mbox{card}({\bf Z};{\bf X})$ the number of the components of ${\bf Z}$ which are $X_i$, $1\leq i\leq d$.
Since $\mu$ is $d$-monotone, then by ($\ref{3002}$) we have that
\begin{align*}
 \Delta_{\bf X}^{\bf Y} \Gamma_{\mu}
     & =  \  \underset{{\bf Z}}{\sum} (-1)^{\mbox{card}({\bf Z}; {\bf X})}
             \int_{[0,+\infty)^d} x_1 \cdots x_d L_{\mu,{\bf Z}}(dx_1, \cdots, dx_d) \\
     & =  \  \underset{{\bf Z}}{\sum} (-1)^{\mbox{card}({\bf Z}; {\bf X})}
             \int_0^{+\infty} \cdots \int_0^{+\infty} \mu ( \{ Z_1 > x_1 \} \times \cdots \times \{ Z_d > x_d \} ) dx_1 \cdots dx_d \\
     & =  \  \int_0^{+\infty} \cdots \int_0^{+\infty} \underset{{\bf Z}}{\sum} (-1)^{\mbox{card}({\bf Z}; {\bf X})}
               \mu ( \{ Z_1 > x_1 \} \times \cdots \times \{ Z_d > x_d \} ) dx_1 \cdots dx_d \\
     & =  \  \int_0^{+\infty}\cdots\int_0^{+\infty} \Delta_{\{X_1>x_1\}\times \cdots \times\{ X_d >x_d\}}^{\{Y_1 >x_1\} \times \cdots \times \{Y_d >x_d\}}
               \mu dx_1 \cdots dx_d \\
     & \geq  0,
\end{align*}
where the summation $\underset{{\bf Z}}{\sum}$ is summed over the $2^d$  random vectors ${\bf Z} = (Z_1, \cdots, Z_d).$
The proof of Theorem 3.1 is completed.

\vspace{0.3cm}

\noindent{\bf Proof of Theorem 3.2}

\vspace{0.2cm}

Define a normalized and monotone set function $\mu : \sS \rightarrow \mathbb{R}_+$ by
\begin{align*}
\mu (A_1 \times \cdots \times A_d) := \Gamma \left( (1_{A_1},\cdots,1_{A_d}) \right), \quad A_1 \times \cdots \times A_d \in \sS.
\end{align*}
Since $\Gamma$ is normalized and monotone, hence $\mu$ is normalized, monotone and $\mu_C (\emptyset)= 0$.
Since $\Gamma$ is continuous from below, hence $\mu $ is continuous from below.

\vspace{0.2cm}

Now, we show that $\mu $ is $d$-monotone.
In fact, for any ${\bf A}:= A_1\times \cdots \times A_d$, ${\bf B}:= B_1\times \cdots \times B_d \in \sS$ with $A_i \subseteq B_i$, $1\leq i\leq d$,
write $1_{\bf A} := (1_{A_1},\cdots,1_{A_d})$ and $1_{\bf B}:=(1_{B_1},\cdots,1_{B_d})$.
Then $\Delta_{A_1\times \cdots \times A_d}^{B_1\times \cdots \times B_d}$ $\mu  = \Delta_{1_{\bf A}}^{1_{\bf B}} \Gamma.$
Thus the $d$-monotonicity of $\Gamma$ implies that $\mu $ is $d$-monotone.

\vspace{0.2cm}

Next, we proceed to show ($\ref{3003}$). Given any ${\bf X}= (X_1,\cdots,X_d)\in \sX^d_+$,
for each $1\leq i\leq d$, let $n_i$ be a positive integer, and define
\begin{align*}
  u_{n_i,i}(x):= \frac{1}{2^{n_i}}\sum_{j_{i}=1}^{n_{i} \cdot 2^{n_{i}}-1}{{1}}_{\{x>\frac{j_i}{2^{n_i}}\}}, \quad x \geq 0.
 \end{align*}

Note that for each $1\leq i\leq d$, $j_i \in \{ 1, \cdots, n_i 2^{n_i}-1 \}$ and any $x\geq 0$,
\begin{align}\label{4007}
  2^{n_i} \cdot \left( x\wedge\frac{j_i+1}{2^{n_i}}-x\wedge\frac{j_i}{2^{n_i}} \right) \leq {{1}}_{\{x>\frac{j_i}{2^{n_i}}\}} \leq 2^{n_i} \cdot
       \left( x\wedge\frac{j_i}{2^{n_i}}-x\wedge\frac{j_i-1}{2^{n_i}} \right),
\end{align}

\begin{align}\label{4010}
  \sum_{j_i=1}^{n_i\cdot 2^{n_i}-1}\left( X_i\wedge \frac{j_i+1}{2^{n_i}}- X_i\wedge \frac{j_i}{2^{n_i}} \right)
     = X_i\wedge n_i- X_i\wedge \frac{1}{2^{n_i}}
\end{align}
and
\begin{align}\label{4011}
  \sum_{j_i=1}^{n_i\cdot 2^{n_i}-1}\left( X_i\wedge \frac{j_i}{2^{n_i}}- X_i\wedge \frac{j_i-1}{2^{n_i}} \right)
           = X_i\wedge \left( n_i- \frac{1}{2^{n_i}} \right)
           \leq X_i.
\end{align}
Hence, when $n_i$ goes to infinity, $u_{n_i,i}(X_i)$ almost surely converges to $X_i$.

\vspace{0.2cm}

From (\ref{4007}), the definition of $\mu$ and the monotonicity of $\Gamma$,
it follows that for each $1\leq i\leq d$ and $j_i$ $\in$ $\{ 1, \cdots, n_i 2^{n_i}-1 \}$,
\begin{align*}
\Gamma & \left(\left( 2^{n_1} \cdot \left(X_1 \wedge\frac{j_1+1}{2^{n_1}}-X_1\wedge\frac{j_1}{2^{n_1}}\right), \cdots, 2^{n_d} \cdot
          \left(X_d \wedge\frac{j_d+1}{2^{n_d}}-X_1\wedge\frac{j_d}{2^{n_d}}\right)\right) \right)\\
       & \leq \mu \left( \left \{ X_1>\frac{j_1}{2^{n_1}} \right \} \times \cdots \times \left \{ X_d >\frac{j_d}{2^{n_d}} \right \} \right)\\
       & \leq  \Gamma\left( \left( 2^{n_1} \cdot \left(X_1 \wedge\frac{j_1}{2^{n_1}}-X_1\wedge\frac{j_1-1}{2^{n_1}}\right),\cdots,
              2^{n_d} \cdot \left(X_d\wedge\frac{j_d}{2^{n_d}}-X_d\wedge\frac{j_d-1}{2^{n_d}} \right) \right) \right).
\end{align*}
Since $\Gamma$ is component-wise positively homogeneous, thus
\begin{align}\label{241108add1}
  & 2^{n_1+\cdots+n_d} \cdot \Gamma\left(\left(X_1 \wedge\frac{j_1+1}{2^{n_1}}-X_1\wedge\frac{j_1}{2^{n_1}}, \cdots ,X_d
                          \wedge\frac{j_d+1}{2^{n_d}}-X_1\wedge\frac{j_d}{2^{n_d}}\right) \right) \nonumber\\
          & \leq \mu \left( \left\{ X_1>\frac{j_1}{2^{n_1}} \right\} \times \cdots \times \left\{ X_d>\frac{j_d}{2^{n_d}} \right\} \right) \nonumber\\
          & \leq  2^{n_1+\cdots+n_d} \cdot \Gamma\left(\left(X_1 \wedge\frac{j_1}{2^{n_1}}-X_1\wedge\frac{j_1-1}{2^{n_1}},\cdots,
                     X_d\wedge\frac{j_d}{2^{n_d}}-X_d\wedge\frac{j_d-1}{2^{n_d}}\right) \right).
\end{align}
By Lemma 2.2(1), for any ${\bf n}:= (n_1,\cdots,n_d) \in \mathbb{N}^d$, any two elements from the set
\begin{align*}
  & \left \{ \left( \sum_{j_1=1}^{k_1} \left( X_1 \wedge \frac{j_1+1}{2^{n_1}} - X_1 \wedge \frac{j_1}{2^{n_1}} \right), \cdots,
                       \sum_{j_d=1}^{k_d} \left( X_d \wedge \frac{j_d+1}{2^{n_d}} - X_d \wedge \frac{j_d}{2^{n_d}} \right) \right); \right. \\
  & \quad \quad  \left. \ k_i \in \{ 1, \cdots, n_i \cdot 2^{n_i} \}, 1\leq i\leq d
 \right \}
\end{align*}
are comonotone. By Lemma 2.2(2), for any given $(k_1, \cdots, k_d)$, $k_i \in \{ 1,\cdots,n_i\cdot 2^{n_i} \}$, $1\leq i\leq d$, the random vector $(\sum_{j_1=1}^{k_1} ( X_1 \wedge \frac{j_1+1}{2^{n_1}} - X_1 \wedge \frac{j_1}{2^{n_1}} ), \cdots, \sum_{j_d=1}^{k_d} ( X_d \wedge \frac{j_d+1}{2^{n_d}}
- X_d \wedge \frac{j_d}{2^{n_d}}))$ is comonotone with $( Y_1, \cdots, Y_d )$, where one of the component of $(Y_1,\cdots,Y_d)$, say $Y_l$,
equals to $( X_l \wedge \frac{k_l+2}{2^{n_l}} - X_l \wedge \frac{k_l+1}{2^{n_l}} )$, and the others $Y_i$, $i\neq l$, equal to
$\sum_{j_i=1}^{k_i} ( X_i \wedge \frac{j_i+1}{2^{n_i}} - X_i \wedge \frac{j_i}{2^{n_i}} )$.
Hence, by (\ref{241108add1}) and the comonotone additivity of $\Gamma,$  we know that
\begin{align*}
\Gamma & \left(\left(\sum_{j_1=1}^{n_1\cdot 2^{n_1}-1}\left(X_1\wedge \frac{j_1+1}{2^{n_1}}-X_1\wedge \frac{j_1}{2^{n_1}}\right), \cdots,
              \sum_{j_d=1}^{{n_d}\cdot 2^{n_d}-1}\left(X_d\wedge \frac{j_d+1}{2^{n_d}}-X_d\wedge \frac{j_d}{2^{n_d}}\right)\right) \right) \\
   & \leq \frac{1}{2^{n_1+\cdots+n_d}} \cdot \sum_{j_1=1}^{{n_1}\cdot 2^{n_1}-1}\cdots \sum_{j_d=1}^{{n_d}\cdot 2^{n_d}-1}
          \mu \left( \left\{ X_1>\frac{j_1}{2^{n_1}} \right\} \times \cdots \times \left\{ X_d>\frac{j_d}{2^{n_d}} \right\} \right)\\
   & \leq \Gamma \left( \left( \sum_{j_1=1}^{{n_1}\cdot 2^{n_1}-1}\left( X_1\wedge \frac{j_1}{2^{n_1}}-X_1\wedge \frac{j_1-1}{2^{n_1}} \right), \cdots,
          \sum_{j_d=1}^{{n_d}\cdot 2^{n_d}-1}\left( X_d\wedge \frac{j_d}{2^{n_d}}-X_d\wedge \frac{j_d-1}{2^{n_d}} \right) \right) \right),
\end{align*}
which, together with (\ref{4010}), (\ref{4011}) and the monotonicity of $\Gamma$, yields that
\begin{align}\label{4003}
\Gamma & \left( \left(  X_1\wedge n_1-X_1\wedge \frac{1}{2^{n_1}}, \cdots, X_d \wedge n_d-X_d\wedge \frac{1}{2^{n_d}}\right) \right)\nonumber\\
       & \quad \leq \frac{1}{2^{n_1+\cdots+n_d}} \cdot \sum_{j_1=1}^{n_1\cdot 2^{n_1}-1} \cdots \sum_{j_d=1}^{n_d\cdot 2^{n_d}-1}
                 \mu \left( \left\{ X_1>\frac{j_1}{2^{n_1}} \right\} \times \cdots \times \left\{ X_d>\frac{j_d}{2^{n_d}} \right\} \right)\nonumber\\
       & \quad \leq \Gamma \left( \left( X_1\wedge\left( n_1-\frac{1}{2^{n_1}} \right), \cdots,
                            X_d\wedge\left(n_d-\frac{1}{2^{n_d}}\right)\right), \right)\nonumber\\
       & \quad \leq \Gamma({\bf X}).
\end{align}
Consequently, by (\ref{4003}) and the continuity from below of $\Gamma$, we have that
\begin{align}\label{4004}
  & \Gamma({\bf X})\notag\\
  & = \underset{n_d\rightarrow +\infty}{\lim} \cdots \underset{n_1 \rightarrow +\infty}{\lim} \frac{1}{2^{n_1+\cdots+n_d}} \cdot
        \sum_{j_1=1}^{n_1\cdot 2^{n_1}-1} \cdots \sum_{j_d=1}^{n_d\cdot 2^{n_d}-1}
                   \mu \left( \left\{ X_1>\frac{j_1}{2^{n_1}} \right\} \times \cdots \times \left\{ X_d>\frac{j_d}{2^{n_d}} \right\} \right).
\end{align}

\vspace{0.2cm}

Now, we turn to calculate the right-hand side of (\ref{4004}). First, we claim that
\begin{align}\label{4005}
   & \frac{1}{2^{n_1+\cdots+n_d}} \cdot \sum_{j_1=1}^{n_1\cdot 2^{n_1}-1} \cdots \sum_{j_d=1}^{n_d\cdot 2^{n_d}-1}
                \mu \left( \left\{ X_1>\frac{j_1}{2^{n_1}} \right\} \times \cdots \times \left\{ X_d>\frac{j_d}{2^{n_d}} \right\} \right)\nonumber\\
   & \qquad  = \int_0^{+\infty} \cdots \int_0^{+\infty}
                  \mu \left( \left\{ u_{n_1,1}(X_1)>x_1 \right\} \times \cdots \times \left\{ u_{n_d,d}(X_d)>x_d \right\} \right) dx_1\cdots dx_d.
\end{align}
In fact, for any $(x_1, \cdots, x_d) \in [ \frac{j_1}{2^{n_1}}, \frac{j_1+1}{2^{n_1}} ) \times \cdots
    \times [ \frac{j_d}{2^{n_d}}, \frac{j_d+1}{2^{n_d}} ),$ $ j_i \in \{0, 1, \cdots, n_i 2^{n_i}-2 \}$, $1\leq i\leq d,$
\begin{align*}
  \mu & \left( \left\{ u_{n_1,1}(X_1)>x_1 \right\} \times \cdots \times \left\{ u_{n_d,d}(X_d)>x_d \right\}  \right) \\
        & \quad = \mu \left( \left\{ u_{n_1,1}(X_1)>\frac{j_1}{2^{n_1}} \right\} \times \cdots \times \left\{ u_{n_d,d}(X_d)>\frac{j_d}{2^{n_d}}
                                                                                                               \right\} \right) \\
        & \quad = \mu \left( \left\{ X_1>\frac{j_1+1}{2^{n_1}} \right\} \times \cdots \times \left\{ X_d>\frac{j_d+1}{2^{n_d}} \right\} \right),
\end{align*}
while, if there exists some $i \in \{1,\cdots,d\}$ such that $x_i\geq \frac{n_i\cdot 2^{n_i}-1}{2^{n_i}}$, then
\begin{align*}
  \mu \left( \left\{ u_{n_1,1}(X_1)>x_1 \right\} \times \cdots \times \left\{ u_{n_d,d}(X_d)>x_d \right\} \right) = \mu_C (\emptyset) = 0.
\end{align*}
Therefore, (\ref{4005}) holds.

\vspace{0.2cm}

Second, from the monotonicity of $\mu $, (\ref{4007}), (\ref{4010}) and (\ref{4011}), it follows that for any $(x_1,\cdots,x_d)\in [0,+\infty)^d$,
\begin{align}\label{4006}
  \mu & \left( \left\{ X_1\wedge n_1- X_1 \wedge \frac{1}{2^{n_1}}>x_1 \right\} \times \cdots
                             \times \left\{ X_d\wedge n_d- X_d \wedge \frac{1}{2^{n_d}}>x_d \right\} \right) \nonumber\\
        & \quad \leq \mu \left( \left\{ u_{n_1,1}(X_1)>x_1 \right\} \times \cdots \times \left\{ u_{n_d,d}(X_d)>x_d \right\} \right) \nonumber\\
        & \quad \leq \mu \left( \left\{ X_1>x_1 \right\} \times \cdots \times \left\{ X_d>x_d \right\} \right).
\end{align}

\vspace{0.2cm}

Note that for each $1\leq i\leq d$, when $n_i$ goes to infinity, the random variables $X_i\wedge n_i- X_i\wedge \frac{1}{2^{n_i}}$ increase
almost surely to the random variable $X_i$. Therefore, from (\ref{4006}), the monotone convergence theorem and the continuity from below of $\mu,$
it follows that
\begin{align} \label{4008}
 \underset{n_d\rightarrow +\infty}{\lim}
   & \cdots \underset{n_1 \rightarrow +\infty}{\lim} \int_0^{\infty}\cdots \int_0^{\infty}
        \mu \left( \left\{ u_{n_1,1}(X_1)>x_1 \right\} \times \cdots \times \left\{ u_{n_d,d}(X_d)>x_d \right\} \right) dx_1 \cdots dx_d \nonumber\\
   & = \int_0^{\infty} \cdots \int_0^{\infty} \mu \left( \left\{ X_1>x_1 \right\} \times \cdots \times \left\{ X_d>x_d \right\} \right) dx_1 \cdots dx_d.
\end{align}
Consequently, ($\ref{3003}$) follows from (\ref{4004}), (\ref{4005}) and (\ref{4008}). Theorem 3.2 is proved.

\vspace{0.3cm}

\noindent{\bf Proof of Theorem 3.3}

\vspace{0.2cm}

Let $\mu$ be the normalized and $d$-monotone set function on $\sS$ as in Theorem 3.2 such that (\ref{3003}) holds, that is,
for any ${\bf W} = (W_1, \cdots, W_d) \in \sX^d_+,$
\begin{align}\label{4002}
  \Gamma \left( {\bf W} \right) = \int_0^{+\infty}\cdots \int_0^{+\infty}
         \mu ( \{ W_1 > t_1 \} \times \cdots \times \{ W_d > t_d \} ) dt_1 \cdots dt_d,
\end{align}
where $\mu (A_1 \times \cdots \times A_d) := \Gamma \left((1_{A_1}, \cdots, 1_{A_d}) \right)$
for $A_i \in \sF$, $1\leq i \leq d$.

\vspace{0.2cm}

Let ${\bf X} = (X_1, \cdots, X_d) \in \sX^d_+$ be a given continuous random vector, which has a unique copula $C.$
Recall that $F_{X_i}(x_i) := P(X_i \leq x_i)$ and
$S_{X_i}(x_i) := P(X_i > x_i), x_i \in \mathbb{R},$ are the distribution function and survival function of $X_i$ under probability measure $P$,
respectively, $1\leq i \leq d.$
For each $1\leq i \leq d,$ define $U_i := F_{X_i}(X_i)$, and define ${\bf U}$ $ := $ $(U_1, \cdots, U_d)$ $\in$ $ \sX^d_+.$
Clearly, for each $1\leq i \leq d,$ $0 \leq U_i(\omega) \leq 1$ for every $\omega \in \Omega,$ and $U_i$ is uniformly distributed on $(0,1),$
for instance, see F\"{o}llmer and Schied (2016, Lemma A.25). By Lemma 2.1(1), we know that  $C$ is a copula for ${\bf U}$.

\vspace{0.2cm}

Define a function $\gamma_{_C} : \ [0,1]^d \rightarrow [0, 1]$ by
\begin{align}\label{221102add1}
  \gamma_{_C}(u_1, \cdots, u_d)
       := \Gamma \left((1_{\{ U_1 > 1-u_1 \}}, \cdots, 1_{\{ U_d > 1-u_d \}}) \right), \quad (u_1, \cdots, u_d) \in [0,1]^d.
\end{align}
Notice that $U_i := F_{X_i}(X_i)$ is uniformly distributed on $(0,1)$ for any continuous random variable $X_i$. Hence,
by the Sklar's Theorem and the distribution invariance of $\Gamma$,
we know that the right-hand side of (\ref{221102add1}) is determined by the copula for the random vector
$(1_{\{ U_1 > 1-u_1 \}}, $ $\cdots,$ $ 1_{\{ U_d > 1-u_d \}})$, and that $C$ is a copula for it due to Lemma 2.1(2).
Thus, we have adopted the copula $C$ rather than ${\bf X}$ itself as a subscript in the notation  $\gamma_{_C}$ to emphasize its dependence
on the copula $C$.
In the sequel, we will find that the desired distortion functions $g_{_{1,C}}, \cdots, g_{_{d,C}}$ also depend on the copula $C.$

\vspace{0.2cm}

We first claim that $\gamma_{_C}$ defined as in (\ref{221102add1}) is a grounded $d$-increasing function in the sense of
Nelsen (2006, Definition 2.10.2 and page 44).
In fact, the normalization and  monotonicity of $\Gamma$ imply that
$ 0  \leq  \gamma_{_C}(u_1, \cdots, u_d) \leq $ $\Gamma({\bf 1}) = 1$ for any $(u_1,\cdots,u_d)\in [0,1]^d$.
For any $ 0 \leq v_i \leq t_i \leq 1,$ $ i=1, \cdots, d$, from the $d$-monotonicity of $\Gamma$, it follows that
\begin{align}\label{4009}
\Delta_{(v_1,\cdots,v_d)}^{(t_1,\cdots,t_d)} \gamma_{_C}
  & = \Delta_{v_d}^{t_d} \cdots \Delta_{v_1}^{t_1} \Gamma\left( (1_{\{ U_1 > 1-  \cdot \}}, \cdots, 1_{\{ U_d > 1- \cdot \}}) \right) \nonumber \\
  & = \Delta_{(1_{\{ U_1 > 1-v_1 \}},\cdots,1_{\{ U_d > 1-v_d \}})}^{(1_{\{ U_1 > 1-t_1 \}}, \cdots, 1_{\{ U_d > 1-t_d \}})} \Gamma \nonumber\\
  & \geq 0,
\end{align}
which yields that $\gamma_{_C}$ is $d$-increasing.
Clearly, for any $(u_1, \cdots, u_d) \in [0,1]^d,$ whenever there is at least one $i$ such that $u_i = 0,$ then
\begin{align*}
      & \gamma_{_C}(u_1, \cdots, u_{i-1}, 0, u_{i+1}, \cdots, u_d) \\
      & \quad = \Gamma ( ( 1_{\{ U_1 > 1-u_1 \}}, \cdots, 1_{\{ U_{i-1} > 1-u_{i-1} \}}, 1_{\emptyset}, 1_{\{ U_{i+1} > 1-u_{i+1} \}},
                                                                                                            \cdots, 1_{\{ U_d > 1-u_d \}}))
         =  0,
\end{align*}
which exactly means that $\gamma_{_C}$ is grounded.
In summary, $\gamma_{_C}$ is a grounded $d$-increasing function in the sense of Nelsen (2006).
Moreover, from the continuity from below of $\Gamma$, it follows that $\gamma_{_C}$ is left-continuous with respect to each variable on $(0,1].$

\vspace{0.2cm}

Furthermore, notice that the two random variables $ 1_{\{ U_i > 0 \}} $ and $1_\Omega$ have the same distribution function, $1\leq i \leq d.$
By Lemma 2.1(2) we know that $C$ is a copula for both random vectors $( 1_{\{ U_1 > 0 \}}, \cdots,  1_{\{ U_d > 0 \}})$ and $(1_\Omega, \cdots, 1_\Omega)$.
Hence,
$( 1_{\{ U_1 > 0 \}}, \cdots,  1_{\{ U_d > 0 \}})$ and $(1_\Omega, \cdots, 1_\Omega)$ have the same joint distribution function.
Thus, by the normalization and  distribution invariance of $\Gamma,$ we have that
\begin{align}\label{221102add2}
  \Gamma ( ( 1_{\{ U_1 > 0 \}}, \cdots,  1_{\{ U_d > 0 \}})) = \Gamma ((1_\Omega, \cdots, 1_\Omega)) = 1.
\end{align}

\vspace{0.2cm}

For each $1\leq i \leq d,$ the $i$th one-dimensional margin of $\gamma_{_C},$ denoted by $g_{_{i,C}},$ is defined by
\begin{align}\label{221102add4}
g_{_{i,C}}(u_i)  := \gamma_{_C}(1, \cdots, 1, u_i, 1, \cdots, 1), \quad u_i \in [0, 1],
\end{align}
for instance, see Nelsen (2006, (2.10.2) on page 44).
Moreover, notice that $g_{_{i,C}}$ is a distortion function, and is left-continuous on $(0, 1].$
Indeed, the left-continuity and monotonicity of $g_{_{i,C}}$ are straightforward, due to the continuity from below
and monotonicity of $\Gamma,$ respectively.
By (\ref{221102add2}) and (\ref{221102add4}),
\begin{align*}
g_{_{i,C}}(1)  = \Gamma ( ( 1_{\{ U_1 > 0 \}}, \cdots,  1_{\{ U_d > 0 \}})) = 1.
\end{align*}
Apparently,
\begin{align*}
g_{_{i,C}}(0) = \Gamma ( ( 1_{\{ U_1 > 0 \}}, \cdots, 1_{\{ U_{i-1}> 0 \}}, 1_\emptyset, 1_{\{ U_{i+1} > 0 \}}, \cdots,  1_{\{ U_d > 0 \}}))
           = 0.
\end{align*}

\vspace{0.2cm}

Define a function $C^*(u_1, \cdots, u_d): \ \mbox{Ran}(g_{_{1,C}}) \times \cdots \times \mbox{Ran}(g_{_{d,C}}) \longrightarrow [0, 1]$ by
\begin{align}\label{221103add7}
   C^*(u_1, \cdots, u_d):=\gamma_{_C}(g_{_{1,C}}^{-1}(u_1), \cdots, g_{_{d,C}}^{-1}(u_d)),
\end{align}
where $g_{_{i,C}}^{-1}$ is the right-continuous inverse function of $g_{_{i,C}},$ that is,
\begin{align*}
g_{_{i,C}}^{-1}(u_i):=
  \begin{cases}
              \inf \{ x \in [0,1] : \ g_{_{i,C}}(x) > u_i \}, & \mbox{if } 0 < u_i < 1, \\
               0,                                          & \mbox{if } u_i=0, \\
               1,                                          & \mbox{if } u_i=1.
  \end{cases}
\end{align*}
We conclude that $C^*$ is a sub-copula. In fact, for each $1\leq i\leq d$,
\begin{align*}
 C^* & (u_1, \cdots, u_{i-1}, 0, u_{i+1}, \cdots, u_d)\\
  & := \gamma_{_C}(g_{_{1,C}}^{-1}(u_1), \cdots, g_{_{{i-1},C}}^{-1}(u_{i-1}), g_{_{i,C}}^{-1}(0), g_{_{{i+1},C}}^{-1}(u_{i+1}), \cdots, {g_{_{d,C}}}^{-1}(u_d))\\
  & = \gamma_{_C}(g_{_{1,C}}^{-1}(u_1), \cdots, g_{_{{i-1},C}}^{-1}(u_{i-1}), 0, g_{_{{i+1},C}}^{-1}(u_{i+1}), \cdots, g_{_{d,C}}^{-1}(u_d))\\
  & = 0.
\end{align*}
If ${g_{_{1,C}}}, \cdots, {g_{_{d,C}}}$ are continuous, then $\mbox{Ran} (g_{_{i,C}}) = [0, 1], 1\leq i \leq d.$ Hence, by (\ref{221102add4}),
for any $u_i \in [0, 1],$
\begin{align} \label{4001}
 C^* & (1, \cdots, 1, u_i, 1, \cdots, 1)\nonumber \\
     & = \gamma_{_C}(g_{_{1,C}}^{-1}(1), \cdots, g_{_{{i-1},C}}^{-1}(1), g_{_{i,C}}^{-1}(u_i), g_{_{{i+1},C}}^{-1}(1), \cdots, g_{_{d,C}}^{-1}(1))\nonumber\\
     & = \gamma_{_C}(1, \cdots, 1, g_{_{i,C}}^{-1}(u_i), 1, \cdots, 1) \nonumber\\
     & = g_{_{i,C}}(g_{_{i,C}}^{-1}(u_i)) \nonumber\\
     & = u_i.
\end{align}
If for some $i\in \{1,\cdots,d\}$, ${g_{_{i,C}}}$ is not continuous, then (\ref{4001}) is still true for $u_i \in \mbox{Ran} (g_{_{i,C}})$.
Taking (\ref{4009}) into account, we have that for any $0\leq v_i \leq t_i\leq 1,$ $v_i,$ $t_i \in \mbox{Ran} (g_{_{i,C}}),$ $ 1\leq i \leq d,$
\begin{align*}
  \Delta_{(v_1,\cdots,v_d)}^{(t_1,\cdots,t_d)} C^*
     = \Delta_{(\theta_1,\cdots,\theta_d)}^{(\tau_1,\cdots,\tau_d)} \gamma_{_C} \geq 0,
\end{align*}
where $\theta_i:= g_{_{i,C}}^{-1}(v_i)$, $\tau_i:= g_{_{i,C}}^{-1}(t_i)$, $1\leq i \leq d$, and the last inequality is due to (\ref{4009}).
Consequently, $C^*$ is a sub-copula. Particularly, if ${g_{_{1,C}}}, \cdots, {g_{_{d,C}}}$ are continuous, then $C^*$ is a copula. In any case,
by the copula extension lemma (see pages 46-47 and Lemma 2.3.5 of Nelsen (2006)), $C^*$ can be extended to a copula on $[0,1]^d$,
which is still denoted by $C^*$.

\vspace{0.3cm}

Furthermore, as a consequence of Nelsen (2006, Lemma 2.10.4) we know that for any $\alpha_1,$ $\cdots,$ $\alpha_d$ $\in$ $[0,1],$
\begin{align}
  \gamma_{_C} (\alpha_1, \cdots, \alpha_d)
          & = \gamma_{_C}( g_{_{1,C}}^{-1} (g_{_{1,C}}(\alpha_1)), \cdots,  g_{_{d,C}}^{-1} (g_{_{d,C}}(\alpha_d))) \nonumber \\
          & = C^*( g_{_{1,C}}(\alpha_1), \cdots,  g_{_{d,C}}(\alpha_d)), \label{221103add3}
\end{align}
where the second equality is apparent, due to the definition of $C^*.$

\vspace{0.3cm}

Next, we proceed to show that (\ref{3004}) holds for the continuous random vector ${\bf X}.$
We first claim that for any $t_1, \cdots, t_d \in \mathbb{R},$
\begin{align} \label{221103add4}
 \Gamma((1_{\{X_1>t_1\}}, \cdots, 1_{\{X_d>t_d\}})) =  \Gamma((1_{\{F_{X_1}(X_1)>F_{X_1}(t_1)\}}, \cdots, 1_{\{F_{X_d}(X_d)>F_{X_d}(t_d)\}})).
\end{align}
Indeed, by the distribution invariance of $\Gamma,$ it suffices to show that $(1_{\{X_1>t_1\}}, \cdots, 1_{\{X_d>t_d\}})$ and
$ (1_{\{F_{X_1}(X_1)>F_{X_1}(t_1)\}}, \cdots, 1_{\{F_{X_d}(X_d)>F_{X_d}(t_d)\}}) $ have the same joint distribution function under $P$.
Note that by Lemma 2.1, $C$ is a copula for both random vectore $ (1_{\{X_1>t_1\}}, \cdots, 1_{\{X_d>t_d\}}) $ and
$ (1_{\{F_{X_1}(X_1)>F_{X_1}(t_1)\}}, \cdots, 1_{\{F_{X_d}(X_d)>F_{X_d}(t_d)\}})$. Hence,
it is further sufficient for us to show that for each $1\leq i \leq d,$ the random variables $ 1_{\{X_i>t_i\}}$ and $ 1_{\{F_{X_i}(X_i)>F_{X_i}(t_i)\}}$
have the same distribution function, which is equivalent to proving
\begin{align} \label{221103add5}
 P(X_i > t_i) = P(F_{X_i}(X_i)>F_{X_i}(t_i)).
\end{align}
To verify (\ref{221103add5}). Denote $t^*_i := \sup \{ t \in \mathbb{R} : \ F_{{X_i}}(t) \leq F_{{X_i}}(t_i)\},$ then $t^*_i \in [t_i, +\infty]$
and $ F_{{X_i}}(t^*_i) = F_{{X_i}}(t_i).$ Hence
\begin{align*}
P(X_i > t^*_i) = P(X_i > t_i) \quad \mbox{and} \quad \{ F_{X_i}(X_i) > F_{X_i}(t^*_i) \} = \{ F_{X_i}(X_i) > F_{X_i}(t_i) \}.
\end{align*}
Therefore, to show (\ref{221103add5}), it suffices to show
\begin{align} \label{221103add6}
 \{X_i > t^*_i \} = \{ F_{X_i}(X_i)>F_{X_i}(t^*_i) \}.
\end{align}
In fact, without any loss of generality, we assume that $t^*_i < +\infty.$ Otherwise, the two sets are empty set, and thus (\ref{221103add6}) holds.
Suppose that an $\omega \in  \{X_i > t^*_i \}$ is given, that is, $X_i(\omega)>t^*_i.$ By the definition of $t^*_i$, we know that
 $ F_{X_i}(X_i(\omega)) > F_{X_i}(t_i) =  F_{X_i}(t^*_i),$ which exactly means that $ \omega \in  \{ F_{X_i}(X_i)>F_{X_i}(t^*_i) \}.$
Thus, $ \{X_i > t^*_i \} \subseteq  \{ F_{X_i}(X_i)>F_{X_i}(t^*_i) \}.$
To show the converse inclusion. Given an $ \omega \in  \{ F_{X_i}(X_i)>F_{X_i}(t^*_i) \},$ that is, $F_{X_i}(X_i(\omega))>F_{X_i}(t^*_i),$
then from the definition of $t^*_i$ it immediately follows
that $ X_i(\omega)>t^*_i,$ since $F_{X_i}$ is non-decreasing. In summary, (\ref{221103add6}) holds, and consequently (\ref{221103add4}) holds.

\vspace{0.3cm}

Now, we turn to show that (\ref{3004}) holds for ${\bf X} = (X_1, \cdots, X_d).$
By (\ref{221103add3}), (\ref{221103add4}) and the definition of $\gamma_{_C}$
 we obtain that for any $t_1,$ $\cdots,$ $t_d$ $\in$ $[0, +\infty),$
\begin{align*}
 \mu (\{ & X_1> t_1 \} \times \cdots \times \{ X_d> t_d \} ) \\
         & := \Gamma((1_{\{X_1>t_1\}}, \cdots, 1_{\{X_d>t_d\}})) \\
         & = \Gamma((1_{\{F_{X_1}(X_1)>F_{X_1}(t_1)\}}, \cdots, 1_{\{F_{X_d}(X_d)>F_{X_d}(t_d)\}})) \\
         & = \Gamma((1_{\{U_1>1-S_{X_1}(t_1)\}}, \cdots, 1_{\{U_d>1-S_{X_d}(t_d)\}})) \\
         & = \gamma_{_C}(S_{X_1}(t_1),\cdots,S_{X_d}(t_d)) \\
         & = C^*(g_{_{1,C}}(P(X_1>t_1)), \cdots, g_{_{d,C}}(P(X_d>t_d))),
\end{align*}
which, together with (\ref{4002}), yields (\ref{3004}).  Theorem 3.3 is proved.

\vspace{0.3cm}

\noindent{\bf Proof of Theorem 3.4}

\vspace{0.2cm}

Recall that for any ${\bf X} = (X_1, \cdots, X_d) \in \sX^d_+$,  the $i$th component of the column vector $H({\bf X})$
is $\Gamma((1, \cdots, 1, X_i, 1, \cdots, 1)),$ $1 \leq i \leq d.$ Properties (1) and (3) are apparent due to the
component-wise positive homogeneity and the monotonicity of $\Gamma.$

\vspace{0.2cm}

Now, we proceed to show the comonotone additivity of $H$. Given any comonotone ${\bf X} = (X_1, \cdots, X_d)$ and
${\bf Y} = (Y_1, \cdots, Y_d) \in \sX^d_+,$ for each $1 \leq i \leq d,$ the $i$th component of the column vector $H({\bf X} + {\bf Y})$ is
$\Gamma((1, \cdots, 1, X_i + Y_i, 1, \cdots, 1)).$ Recall that for any ${\bf Z} \in \sX^d_+,$
$\Gamma ( {\bf Z}) = 0$ if there is at least one zero-valued component of ${\bf Z}$.
Hence, by the comonotone additivity of $\Gamma,$
\begin{align*}
\Gamma & ((1, \cdots, 1, X_i + Y_i, 1, \cdots, 1)) \\
       & = \Gamma((1, \cdots, 1, X_i, 1, \cdots, 1) + (0, \cdots, 0, Y_i, 0, \cdots, 0)) \\
       & = \Gamma((1, \cdots, 1, X_i, 1, \cdots, 1)) + \Gamma((1, \cdots, 1, Y_i, 1, \cdots, 1)),
\end{align*}
which implies the comonotone additivity of $H.$
For any ${\bf c} = (c_1, \cdots, c_d) \in \mathbb{R}^d_+$, by the component-wise positive homogeneity and the normalization of $\Gamma,$
$\Gamma((1, \cdots, 1, c_i, 1, \cdots, 1)) = c_i, 1\leq i\leq d.$
Hence, the translation invariance of $H$ is a simple consequence of the comonotone additivity of $H$.
Theorem 3.4 is proved.

\subsection{\textbf{Lebesque-Stieltjes measures induced by survival functions}}

In this section, we briefly review some basic facts about the Lebesque-Stieltjes measures induced by survival functions.
For more details, we refer to Denneberg (1994).

\vspace{0.2cm}

For convenience, we recall more notations.  Operations on $\sX^d$ are understood in component-wise sense.
For example, for ${\bf X} = (X_1, \cdots, X_d)$, ${\bf Y}$ $=$ $(Y_1,$ $\cdots,$ $ Y_d)$ $\in \sX^d$,
${\bf X} \leq {\bf Y}$ means $X_i(\omega) \leq Y_i(\omega)$ for every $\omega \in \Omega,$ $1\leq i \leq d$.
${\bf X}+{\bf Y}$ stands for $(X_1+Y_1, \cdots, X_d+Y_d)$.
For ${\bf a} = (a_1, \cdots, a_d) \in \mathbb{R}^d$ and $c \in \mathbb{R}$, ${\bf X} \wedge {\bf a}$ means $(X_1 \wedge a_1, \cdots, X_d \wedge a_d)$,
and $c{\bf X}$ means $(cX_1, \cdots, cX_d)$.  For a sequence of random vectors
${\bf X}_n:= (X_{1,n}, \cdots, X_{d,n}) \in \sX^d$, $n\geq 1$, and a random vector ${\bf X}:= (X_1, \cdots, X_d) \in \sX^d,$
${\bf X}_n \uparrow {\bf X}$ means that the sequence of $\{X_{i,n}(\omega); n \geq 1\}$ increasingly converges to $X_i(\omega)$
for every $\omega \in \Omega,$ $1\leq i\leq d$.

\vspace{0.2cm}

For any ${\bf a}=(a_1, \cdots, a_d), {\bf b}=(b_1, \cdots, b_d)\in \mathbb{R}^d$, the increment of a $d$-variate function
$f : \mathbb{R}^d \rightarrow \mathbb{R}$ from ${\bf a}$ to ${\bf b}$ is defined by
\begin{align*}
 \Delta^{{\bf b}}_{{\bf a}}f := \Delta_{{\bf b}, {\bf a}} f
                             := \Delta^{(d)}_{{b_d}, {a_d}}\Delta^{(d-1)}_{{b_{d-1}}, {a_{d-1}}} \cdots \Delta^{(1)}_{{b_1}, {a_1}}f,
\end{align*}
where for any function $g : \mathbb{R}^d \rightarrow \mathbb{R}$,
\begin{align*}
\Delta^{(i)}_{{b_i}, {a_i}} g(x_1, \cdots, x_d)
        := g(x_1, \cdots, x_{i-1}, b_i, x_{i+1}, \cdots, x_d) - g(x_1, \cdots, x_{i-1}, a_i, x_{i+1}, \cdots, x_d).
\end{align*}
Equivalently,
\begin{align*}
\Delta^{\bf b}_{\bf a} f
   & := f(b_1, \cdots, b_d) - [f(a_1, b_2, \cdots, b_d) + \cdots + f(b_1, \cdots, b_{d-1}, a_d)] \\
   & \quad + [f(a_1, a_2, b_3, \cdots, b_d) + \cdots + f(b_1,  \cdots, b_{d-2}, a_{d-1}, a_d)] \\
   & \quad - \cdots + (-1)^d f(a_1, \cdots, a_d).
\end{align*}
A $d$-variate function $f$ is called $d$-monotone, if $\Delta_{{\bf b}, {\bf a}} f \geq 0$ for any ${\bf a} \leq {\bf b}.$
Given a functional $\Gamma : \sX^d \rightarrow \mathbb{R}$, for ${\bf X} = (X_1, \cdots, X_d) \in \sX^d$, we also write
$\Gamma(X_1, \cdots, X_d)$ for $\Gamma({\bf X})$ if the emphasis is on $X_i, 1 \leq i \leq d.$
Similar to $d$-variate functions, we can define the increment of any functional $\Gamma : \sX^d \rightarrow \mathbb{R}$.
More precisely, for any ${\bf X} = (X_1, \cdots, X_d)$, ${\bf Y} = (Y_1, \cdots, Y_d) \in \sX^d$, the increment of
$\Gamma$ from ${\bf X}$ to ${\bf Y}$ is defined by
\begin{align*}
 \Delta^{{\bf Y}}_{{\bf X}} \Gamma := \Delta_{{\bf Y}, {\bf X}} \Gamma
                                   := \Delta^{(d)}_{Y_d, X_d} \Delta^{(d-1)}_{Y_{d-1}, X_{d-1}} \cdots \Delta^{(1)}_{Y_1, X_1} \Gamma,
\end{align*}
where for any functional $G : \sX^d \rightarrow \mathbb{R}$,
\begin{align*}
\Delta^{(i)}_{Y_i, X_i} G(Z_1, \cdots, Z_d)
      & := G(Z_1, \cdots, Z_{i-1}, Y_i, Z_{i+1}, \cdots, Z_d) \\
      & \quad - G(Z_1, \cdots, Z_{i-1}, X_i, Z_{i+1}, \cdots, Z_d).
\end{align*}

\vspace{0.2cm}

Let $g : \ \mathbb{R}^d \rightarrow \mathbb{R}_+$ be a function, which is decreasing and right-continuous with respect to each variable.
For any ${\bf a}=(a_1,\cdots,a_d), {\bf b}=(b_1,\cdots,b_d) \in \mathbb{R}^d$ with ${\bf a} \leq {\bf b}$, the increment of $g$ on $({\bf a},{\bf b}]$
is then defined by
\begin{align*}
  \Delta_{\bf b}^{\bf a}g :=  \Delta_{b_1}^{a_1}\cdots\Delta_{b_d}^{a_d} g,
\end{align*}
where for any function $h : \ \mathbb{R}^d \rightarrow \mathbb{R}_+$,
\begin{align*}
\Delta^{a_i}_{b_i} h(x_1, \cdots, x_d)
        := h(x_1, \cdots, x_{i-1}, a_i, x_{i+1}, \cdots, x_d) - h(x_1, \cdots, x_{i-1}, b_i, x_{i+1}, \cdots, x_d).
\end{align*}
Equivalently,
\begin{align*}
 \Delta_{\bf b}^{\bf a}g
       := & \ g(a_1,\cdots,a_d)- [g(b_1,a_2,\cdots,a_d) + \cdots + g(a_1,\cdots,a_{d-1},b_d)] \\
         & + [g(b_1,b_2,a_3,\cdots,a_d) + \cdots + g(a_1,\cdots,a_{d-2},b_{d-1},b_d)] \\
         & - \cdots + (-1)^d g(b_1,\cdots,b_d),
\end{align*}
if $({\bf a},{\bf b}]\neq \emptyset$; $\Delta_{\bf b}^{\bf a}g:= 0,$ if $({\bf a},{\bf b}]= \emptyset$.

\vspace{0.2cm}

Assume that $\Delta_{\bf b}^{\bf a} g\geq 0$ for any ${\bf a}\leq {\bf b}$. Denote by $\sS\sR$ the semi-ring
$\{ ({\bf a},{\bf b}] : {\bf a}\leq {\bf b}, \ {\bf a},{\bf b} \in \mathbb{R}^d \}$.
Define a mapping $L_g: \sS\sR \rightarrow [0,+\infty)$ by
\begin{align*}
  L_g(({\bf a},{\bf b}]):= \Delta_{\bf b}^{\bf a} g, \quad  \mbox{ if } \ ({\bf a},{\bf b}]\neq \emptyset;
\end{align*}
$L_g (\emptyset) :=0$. Then it can be steadily verified that $L_g$ is a measure on $\sS\sR$.
Therefore, by the Carath\'{e}odory's Measure Expansion Theorem, $L_g$ can be uniquely extended to a measure on
the Borel algebra $\sB(\mathbb{R}^d)$ on $\mathbb{R}^d$, which is called the Lebesgue-Stieltjes (L-S) measure induced by $g$,
and is also denoted by $L_g$.

\vspace{0.2cm}

Recall that $\sS$ denotes the set of all $d$-dimensional measurable rectangles in the product $\sigma$-algebra
$\sF^d := \sF \times \cdots \times \sF.$
Let $\mu$ be a $d$-monotone and continuous from below set function on $\sS.$
For any random vector ${\bf X}= (X_1,\cdots,X_d)\in {\sX}^d,$ the survival function $S_{\mu,{\bf X}}$ of ${\bf X}$
with respect to $\mu$ is defined by
\begin{align*}
    S_{\mu,{\bf X}}(x_1,\cdots,x_d):= \mu ({\bf X}>{\bf x}):= \mu \left( \{X_1>x_1\} \times \cdots \times \{X_d>x_d\} \right),
\end{align*}
${\bf x}= (x_1,\cdots,x_d)\in \mathbb{R}^d$. Note that for any ${\bf a}= (a_1,\cdots,a_d)$, ${\bf b}= (b_1,\cdots,b_d) \in \mathbb{R}^d$
with ${\bf a}\leq {\bf b}$,
\begin{align*}
  \Delta_{\bf b}^{\bf a} S_{\mu,{\bf X}}
         = \Delta_{\left\{ X_1>b_1 \right\}\times \cdots \times \left\{ X_d>b_d \right\}}^{\left\{ X_1>a_1 \right\}\times
                                                                                                       \cdots \times \left\{ X_d>a_d \right\}} \mu.
\end{align*}
Hence, by the properties of $\mu$, we know that $S_{\mu,{\bf X}}$ is decreasing and right-continuous
with respect to each variable, and satisfies $\Delta_{\bf b}^{\bf a} S_{\mu,{\bf X}} \geq 0$ for any ${\bf a}\leq {\bf b}$.
Therefore, by previous discussion, $S_{\mu,{\bf X}}$ can induce an L-S measure on $\sB(\mathbb{R}^d)$, which is denoted by $L_{\mu,{\bf X}}$.
Particularly, for any ${\bf a} = (a_1, \cdots, a_d),$ $ {\bf b} = (b_1, \cdots, b_d) \in \mathbb{R}^d$ with ${\bf a}\leq{\bf b}$,
\begin{align*}
 L_{\mu, {\bf X}}(({\bf a},{\bf b}]) := \Delta^{\bf a}_{\bf b} S_{\mu, {\bf X}}.
\end{align*}
Moreover, for any ${\bf a} = (a_1, \cdots, a_d) \in \mathbb{R}^d$,
write $({\bf a}, +\infty)$ $:=$ $\{(x_1, \cdots, x_d)$ $\in $ $\mathbb{R}^d :$ $ a_i $ $ < $ $x_i $ $ < $ $ +\infty,$ $ 1 \leq i \leq d \},$
then
\begin{align*}
L_{\mu, {\bf X}}(({\bf a},+\infty))
   = S_{\mu, {\bf X}}(a_1,\cdots,a_d)
   = \mu(\{X_1 > a_1 \} \times \cdots \times \{X_d > a_d \}).
\end{align*}

\subsection{\textbf{ Distortion joint risk measures on $\sX^d$}}

In this section, we address how to extend the main results for distortion joint risk measures from $\sX^d_+$ to $\sX^d$.
In this situation, all the notations and axioms for functionals are automatically parallelly extended from $\sX^d_+$ to $\sX^d.$
For simplicity of presentation, we concentrate on the case of $ d = 2$.

\vspace{0.2cm}

Next theorem extends Theorem 3.2  to general two-dimensional random vectors. By the same arguments,
it would not be hard to extend Theorems 3.3 and 3.4 to the general two-dimensional random vectors ${\bf X} \in \sX^2.$
Nevertheless, due to spaces, we will omit their detailed representations at this moment.

\vspace{0.2cm}

\noindent{\bf Theorem A.1} \ \ Suppose that a normalized joint risk measure $\Gamma: \sX^2  \rightarrow \mathbb{R}$
satisfies the Axioms (A1)-(A5). Then there exists a normalized, $2$-monotone and continuous from below set function $\mu$ on $\sS$
depending on $\Gamma,$ such that for any  ${\bf X}$ $ = $ $ (X_1, X_2)$ $ \in $ $ \sX^2,$
\begin{align}\label{1020add1}
 \Gamma({\bf X})
  & = \int_{-\infty}^0 \left [ \int_{-\infty}^0 \left[ \mu (\{X_1>x_1\} \times \{X_2>x_2\}) - \mu (\Omega \times \{X_2>x_2\}) \right. \right. \nonumber\\
     & \qquad \qquad \qquad     \left. - \mu (\{X_1>x_1\} \times \Omega) + 1 \right] dx_1  \nonumber \\
     & \qquad \qquad \qquad      + \left. \int_0^{\infty} \left[ \mu (\{X_1>x_1\} \times \{X_2>x_2\}) - \mu (\{X_1>x_1\} \times \Omega) \right] dx_1
                                                                                                                                    \right ] dx_2  \nonumber\\
     & \quad + \int_0^{\infty} \left [ \int_{-\infty}^0 \left[ \mu (\{X_1>x_1\} \times \{X_2>x_2\}) - \mu (\Omega \times \{X_2>x_2\}) \right] dx_1 \right.
                                                                                                                                      \nonumber\\
     & \qquad \qquad \qquad      + \left. \int_0^{\infty} \mu(\{X_1>x_1\} \times \{X_2>x_2\}) dx_1 \right ] dx_2.
\end{align}

\vspace{0.2cm}

{\bf Proof} \ \ Let the normalized and monotone set function $\mu :$ \ $\sS $ $\rightarrow$ $\mathbb{R}_+$
be defined as in the proof of Theorem 3.2, that is, for any $A_1 \times A_2 \in \sS,$
\begin{align*}
\mu (A_1 \times A_2) := \Gamma \left( 1_{A_1}, 1_{A_2} \right).
\end{align*}
Then,  $\mu $ is $2$-monotone and continuous from below.

\vspace{0.2cm}

Next, we proceed to show (\ref{1020add1}). To this end, given arbitrarily an ${\bf X}$ $ = $ $ (X_1, X_2)$ $ \in $ $ \sX^2,$
we will show (\ref{1020add1}) by two steps: First, consider the special case where $X_1 \in \sX$ and $X_2 \in \sX_+.$
Second, consider the general case where ${\bf X}$ $ = $ $ (X_1, X_2)$ $ \in $ $ \sX^2.$ Notice that all the two-dimensional random vectors
involved below are comonotone with each other.

\vspace{0.2cm}

Case one: Assume that $X_1 \in \sX$ and $X_2 \in \sX_+.$ Since $X_1$ is bounded, choose $n_1 > \|X_1\|$ so that $(X_1 + n_1, X_2) \in \sX^2_+$.
By the comonotone additivity of $\Gamma$,
\begin{align*}
\Gamma(X_1 + n_1, X_2) = \Gamma((X_1, X_2) + (n_1, 0))
                            = \Gamma(X_1, X_2) + \Gamma(n_1, X_2).
\end{align*}
Hence, by Theorem 3.2 and change-of-variable, we have that
\begin{align}\label{1021add1}
 \Gamma(X_1, X_2)
    & = \Gamma(X_1 + n_1, X_2) - \Gamma(n_1, X_2) \nonumber \\
    & = \int_0^{\infty} \left [ \int_{-\infty}^{0} \left[ \mu( \{X_1 > x_1\} \times \{X_2 > x_2\})
                                                - \mu( \Omega \times \{X_2 > x_2\}) \right] dx_1 \right.   \nonumber \\
    & \qquad \qquad \quad          +   \left. \int_0^{\infty} \mu( \{X_1 > x_1\} \times \{X_2 > x_2\}) dx_1 \right ] dx_2,
\end{align}
which shows that (\ref{1020add1}) holds for any $(X_1, X_2) \in \sX^2$ with $X_2 \in \sX_+.$

\vspace{0.2cm}

Case two: Assume that $(X_1, X_2) \in \sX^2.$ Since $X_2$ is bounded, choose  $n_2 > \|X_2\|$
so that $ X_2 + n_2 \in \sX_+$. By the comonotone additivity of $\Gamma$,
\begin{align}\label{1021add2}
\Gamma(X_1, X_2 + n_2) = \Gamma((X_1, X_2) + (0, n_2))
                            = \Gamma(X_1, X_2) + \Gamma(X_1, n_2).
\end{align}
By (\ref{1021add1}) (i.e. the conclusion of Case one) and change-of-variable, we know that
\begin{align}\label{1021add3}
\Gamma( & X_1, X_2 + n_2) \nonumber \\
        & = \int_{-n_2}^{\infty} \left [ \int_{-\infty}^{0} \left[ \mu( \{X_1 > x_1\} \times \{X_2 > x_2\})
                                                - \mu( \Omega \times \{X_2 > x_2\}) \right] dx_1 \right ] dx_2   \nonumber \\
        & \quad      + \int_{-n_2}^{\infty} \left [ \int_0^{\infty} \mu( \{X_1 > x_1\} \times \{X_2 > x_2\}) dx_1 \right ] dx_2
\end{align}
and
\begin{align}\label{1021add4}
\Gamma(  X_1, n_2)
         = n_2 \int_{-\infty}^{0} \left [ \mu( \{X_1 > x_1\} \times \Omega) - 1 \right ] dx_1
                + n_2 \int_0^{\infty} \mu( \{X_1 > x_1\} \times \Omega) dx_1.
\end{align}
Hence, by dividing the integral $\int_{-n_2}^{\infty}$ in (\ref{1021add3}) with respect to the dummy variable $x_2$ into
$\int_{-n_2}^{0} + \int_{0}^{\infty},$ from (\ref{1021add2})-(\ref{1021add4}), we have that
\begin{align*}
\Gamma ( & X_1, X_2 ) = \Gamma(X_1, X_2 + n_2) - \Gamma(X_1, n_2) \nonumber \\
         & = \int_{-n_2}^{0} \left [ \int_{-\infty}^{0} \left[ \mu( \{X_1 > x_1\} \times \{X_2 > x_2\})
                                                                         - \mu( \Omega \times \{X_2 > x_2\}) \right] dx_1 \right. \nonumber \\
         & \qquad \qquad \qquad  \left. - \int_{-\infty}^{0} \left[ \mu( \{X_1 > x_1\} \times \Omega) - 1 \right] dx_1 \right] dx_2 \nonumber \\
         & \qquad +  \int_{0}^{\infty} \left[ \int_{-\infty}^{0} \left[ \mu( \{X_1 > x_1\} \times \{X_2 > x_2\})
                                                                         - \mu( \Omega \times \{X_2 > x_2\}) \right] dx_1 \right] dx_2  \nonumber \\
         & \quad  + \int_{-n_2}^{0} \left [ \int_0^{\infty} \mu( \{X_1 > x_1\} \times \{X_2 > x_2\}) dx_1
                                                 - \int_0^{\infty} \mu( \{X_1 > x_1\} \times \Omega) dx_1 \right ] dx_2 \nonumber \\
         & \qquad  + \int_{0}^{\infty} \left[ \int_{0}^{\infty} \mu( \{X_1 > x_1\} \times \{X_2 > x_2\}) \right] dx_2 \nonumber \\
         & = \int_{-\infty}^0 \left [ \int_{-\infty}^0 \left[ \mu (\{X_1>x_1\} \times \{X_2>x_2\}) - \mu (\Omega \times \{X_2>x_2\})
                                                                                                                         \right. \right. \nonumber\\
         & \qquad \qquad \qquad     \left. - \mu (\{X_1>x_1\} \times \Omega) + 1 \right] dx_1  \nonumber \\
         & \qquad \qquad \qquad      + \left. \int_0^{\infty} \left[ \mu (\{X_1>x_1\} \times \{X_2>x_2\}) - \mu (\{X_1>x_1\} \times \Omega) \right] dx_1
                                                                                                                                    \right ] dx_2  \nonumber\\
         & \quad + \int_0^{\infty} \left [ \int_{-\infty}^0 \left[ \mu (\{X_1>x_1\} \times \{X_2>x_2\}) - \mu (\Omega \times \{X_2>x_2\}) \right] dx_1
                                                                                                                                    \right. \nonumber\\
         & \qquad \qquad \qquad      + \left. \int_0^{\infty} \mu(\{X_1>x_1\} \times \{X_2>x_2\}) dx_1 \right ] dx_2,
\end{align*}
which is just (\ref{1020add1}). Theorem A.1 is proved.

\end{document}